\pdfoutput=1
\newcommand*{\ATLASLATEXPATH}{latex/}

\documentclass[cernpreprint,texlive=2011,txfonts,UKenglish]{\ATLASLATEXPATH atlasdoc}

\usepackage[subfigure=true]{\ATLASLATEXPATH atlaspackage}
\usepackage{\ATLASLATEXPATH atlasphysics}

\usepackage{\ATLASLATEXPATH atlasbiblatex}
\usepackage{multirow}
\usepackage{textcomp}

\graphicspath{{figures/},{logos/}}

\newcommand\twxs{23.0}

\newcommand\twstat{\pm 1.3}
\newcommand\twsys{^{+3.2}_{-3.5}}
\newcommand\twlumi{\pm 1.1}
\newcommand\sigexp{6.9}
\newcommand\sigobs{7.7}

\newcommand\vtboval{1.01 \pm 0.10}
\newcommand\vtbolim{0.80}

\newcommand\fidxs{0.85}

\newcommand\fidstat{\pm 0.01}
\newcommand\fidsys{^{+0.06}_{-0.07}}
\newcommand\fidlumi{\pm 0.03}

\usepackage{wtxs8-defs}

\usepackage[utf8]{inputenc}

\AtlasTitle{Measurement of the production cross-section of a single top quark in association with a $W$ boson at 8 TeV with the ATLAS experiment}

\author{The ATLAS Collaboration}

\AtlasRefCode{TOPQ-2012-20}

\PreprintIdNumber{CERN-PH-2015-230}

\AtlasJournal{JHEP}
\AtlasAbstract{%
The cross-section for the production of a single top quark in association with 
a $W$ boson in proton--proton collisions at $\sqrt{s}= 8$ TeV is measured. 
The dataset corresponds to an integrated luminosity of 20.3 fb$^{-1}$, collected by the
ATLAS detector in 2012 at the Large  Hadron
Collider at CERN. Events containing two leptons
and one central $b$-jet are selected. The $Wt$ signal is separated from
the backgrounds using boosted decision trees, each of which combines a number of
discriminating variables into one classifier. 
Production of $Wt$ events is observed with a significance of $7.7 \sigma$.
The cross-section is extracted in a profile likelihood fit 
to the classifier output distributions. The $Wt$ cross-section, inclusive of
decay modes, is measured to be $23.0 \pm 1.3$(stat.)$^{+3.2}_{-3.5}$(syst.)$\pm 1.1$(lumi.) pb.
The measured cross-section is used to extract a value for the CKM
matrix element $|V_{tb}|$ of $1.01 \pm 0.10$ and a lower 
limit of 0.80 at the 95\% confidence level.
The cross-section for the production of a top quark and a $W$ boson is also measured 
in a fiducial acceptance requiring two leptons with $p_{\textrm{T}}>25$ GeV and $|\eta|<2.5$,
one jet with $p_{\textrm{T}}>20$ GeV and $|\eta|<2.5$, and
$E_{\textrm{T}}^{\textrm{miss}}>20$ GeV, including both $Wt$ and top-quark pair events as signal.
The measured value of the fiducial cross-section is
$0.85 \pm 0.01$(stat.)$^{+0.06}_{-0.07}$(syst.)$\pm 0.03$(lumi.) pb.
}

\hypersetup{pdftitle={ATLAS draft},pdfauthor={The ATLAS Collaboration}}

\begin{document}

\maketitle
\tableofcontents

%-------------------------------------------------------------------------------
\section{Introduction}
\label{sec:intro}
%-------------------------------------------------------------------------------
The production of a single top quark at the Large Hadron Collider (LHC) proceeds 
via the weak interaction in the Standard Model (SM). The three main modes of single 
top-quark production are: $t$-channel, the exchange of a $W$ boson between a light quark 
and a heavy quark; $s$-channel, via a virtual $W$~boson;
and $Wt$, the production of a top quark in association with a $W$~boson.
Single top-quark production depends on the top-quark coupling to the
$W$~boson, which is parameterised by the form factor $f_{\textrm{LV}}$ and 
the Cabibbo--Kobayashi--Maskawa (CKM) matrix element $V_{tb}$~\cite{CKM1,CKM2,Kane:1991bg}.
The cross-section for each of the three production modes is proportional to the
square of $|f_{\textrm{LV}}V_{tb}|$~\cite{Abazov:2012uga,Alwall:2006bx}.
Physics beyond the SM can contribute to the single top-quark final state and modify the
production cross-sections~\cite{Tait1:2000sh,Cao:2007ea} as well as the kinematic distributions, 
for example through a resonance that decays to $Wt$~\cite{Nutter:2012an,TOPQ-2012-09}.

The production of single top quarks has been observed at the Tevatron
proton--antiproton collider in the $t$-channel~\cite{D0:1p96TeV-tchannel,Aaltonen:2015cra} 
and $s$-channel~\cite{Abazov:2013qka,Aaltonen:2014qja,CDF:OBS-schannel}, as well as their 
combination~\cite{CDF:OBS-SGTOP,D0:OBS-SGTOP,CDF-D0:COMBI-SGTOP}.
The $Wt$ process has a small expected cross-section at the Tevatron and was not observed. 
The $t$-channel mode has been observed by both the ATLAS~\cite{TOPQ-2011-14,TOPQ-2012-21} and
CMS~\cite{CMS-TOP-11-021,CMS-TOP-12-038} collaborations at the LHC. 
The $s$-channel mode has not yet been
measured at the LHC because of its small production cross-section~\cite{TOPQ-2014-03}.
Evidence for $Wt$ production was reported by ATLAS~\cite{TOPQ-2011-17} and
CMS~\cite{CMS-TOP-11-022} in proton--proton ($pp$) collisions at 7~\TeV.
The observation of $Wt$ production 
in $pp$ collisions at 8~\TeV\,has been reported by CMS~\cite{CMS-TOP-12-040}.

\begin{figure}[htbp]
  \begin{center}
\includegraphics[width=0.45\textwidth]{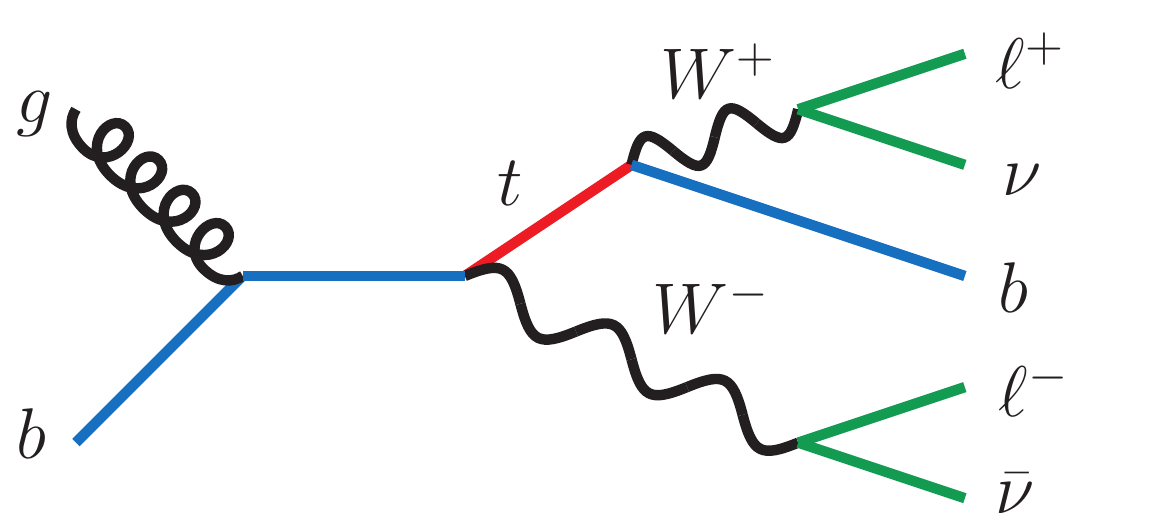}
\caption{Representative leading-order Feynman diagram for the production and decay
of a single top quark in association with a $W$ boson. 
\label{fig:feyn}
}
  \end{center}
\end{figure}

Production of $Wt$ events proceeds via $b$-quark-induced partonic channels such as $gb \rightarrow Wt \rightarrow W^- W^+ b$. A leading-order (LO) Feynman diagram in the
5-flavour-number scheme (5FNS, considering the quarks $u$, $d$, $s$, $c$, and $b$ in the
initial state) is shown in Figure~\ref{fig:feyn}. 
The presence of only a single $b$-quark in the final state represents a distinctive
feature with respect to the $W^+ W^-b \bar{b}$ final state of top-quark pair ($\ttbar$)
production.
The $Wt$ final state contains an additional $b$-quark in higher-order Quantum
Chromodynamics (QCD)
correction diagrams in the 5FNS, as well as in the leading-order process in the
4-flavour-number scheme (4FNS, considering only the quarks $u$, $d$, $s$, $c$ in the
initial state), making it challenging
to experimentally separate $Wt$ production from $\ttbar$ production.

The theoretical prediction for the $Wt$~production cross-section at next-to-leading order (NLO) with 
next-to-next-to-leading logarithmic (NNLL) soft gluon corrections is $22.37 \pm 1.52$~pb~\cite{Kidonakis:2010ux} at a centre-of-mass energy of $\sqrt{s}=$8~\TeV\,for a top-quark mass of
$m_t=172.5$~\GeV~\cite{ATLAS:2014wva}. 
In this calculation, the uncertainty on the theoretical cross-section accounts for the variation of the
renormalisation and factorisation scale between $m_t/2$ and $2 m_t$  and for the
parton distribution function (PDF) uncertainties (using the 90\% confidence level errors of
the MSTW2008 NNLO PDF set~\cite{Martin:2009iq}).
This cross-section represents about 20\% of the total cross-section
for all single top-quark production modes at the LHC.
A second theoretical prediction for the $Wt$~production cross-section is
$18.8 \pm 0.8$~(scale)~$\pm 1.7$~(PDF)~pb, computed at NLO with Hathor 
v2.1~\cite{Aliev:2010zk,Kant:2014oha}. 
The PDF uncertainties are calculated using the
PDF4LHC prescription~\cite{Botje:2011sn} with three different PDF sets ({\scshape CT10},
{\scshape MSTW2008nlo68cl}~\cite{Martin:2009iq} and
{\scshape NNPDF2.3}~\cite{Ball:2013hta}).
The renormalisation and factorisation scales are set to 65~\GeV\,and
the $b$-quark from initial-state radiation is required to have a transverse momentum
of less than 60~\GeV.

This paper presents a measurement of the cross-section for $Wt$~production in $pp$ collisions 
at $\sqrt{s}=$8~\TeV, based on the analysis of \SI[parse-numbers=false]{20.3}{fb^{-1}} of data 
collected by the ATLAS detector in 2012. 
The measurement is carried out in the dilepton final state shown in Figure~\ref{fig:feyn}
where each $W$ boson decays to an electron or a muon and a neutrino ($e\nu$ or $\mu\nu$).
This analysis requires two opposite-sign high-transverse-momentum ($\pt$) leptons
($ee$, $e\mu$, $\mu\mu$), missing transverse momentum ($\etmiss$),
and one high-$\pt$ central jet, which is required to contain a $b$-hadron ($b$-jet).
The main background to this signature is from $\ttbar$
production, with smaller backgrounds coming from dibosons ($WW$, $WZ$, $ZZ$),
$Z$+jets, and events where one or both leptons are misidentified (fake-lepton events) or non-prompt.
Control regions enriched in $\ttbar$ and other background events are also defined. Events in
the $\ttbar$-enriched regions
fulfil the same lepton and missing transverse momentum requirements,
and have exactly two jets, with one or both of the jets required to be identified
as a $b$-jet.
Events in the other background-enriched regions have one or two jets which are required
to not be identified as $b$-jets.
The backgrounds are estimated with simulation, except the non-prompt or fake-lepton background,
which is estimated from data. Boosted decision trees (BDT) are used to optimise the
discrimination between signal and background~\cite{Friedman:2002:SGB:635939.635941}. 
The cross-section is extracted using a profile likelihood fit of the BDT response. 
The background normalisation and the systematic uncertainties are constrained by
simultaneously analysing phase-space regions with substantial $Wt$ signal contributions
and regions where the $Wt$ contributions are negligible.
The ratio of the measured
cross-section to the theoretical prediction (which assumes $V_{tb}=1$)
is used to extract a value of $|f_{\textrm{LV}}V_{tb}|$.

In the 5FNS,
the $Wt$ single top-quark process overlaps and interferes with $\ttbar$ production
at NLO where diagrams involving two top quarks are part of the real emission corrections to
$Wt$ production~\cite{Campbell:2005bb,Frixione:2005vw}. A calculation in the 4FNS scheme
includes $Wt$ and $\ttbar$ as well as non-top-quark diagrams~\cite{Cascioli:2013wga} and
the interference between $Wt$ and $\ttbar$ enters already at tree level.
A measurement of the cross-section inside a fiducial acceptance, designed to reduce the dependence on the
theory assumptions, is also presented. The fiducial acceptance is defined using physics objects constructed
of stable particles to approximate the $Wt$ detector acceptance.
The cross-section for the sum of $Wt$ and $\ttbar$ production is
measured in this fiducial acceptance.

This paper is organised as follows: Section~\ref{sec:detector} provides a brief overview of
the ATLAS detector and the definition of physics objects.
Section~\ref{sec:DataMC} describes the data and Monte Carlo samples 
used for the analysis. Section~\ref{sec:sel} describes the event selection 
and background estimation. Section~\ref{sec:bdt} presents the procedure defined to discriminate 
the signal from the backgrounds using BDTs.
The dominant systematic uncertainties are discussed in Section~\ref{sec:sys}.
Section~\ref{sec:result} presents the results for the inclusive cross-section measurement and for
$|V_{tb}|$ and discusses the impact of systematic uncertainties.
Section~\ref{sec:fid} defines the fiducial acceptance and presents the fiducial cross-section measurement. 
Finally, a summary is presented in Section~\ref{sec:conclusion}.

%-------------------------------------------------------------------------------
\section{The ATLAS detector and object reconstruction}
\label{sec:detector}
%-------------------------------------------------------------------------------

The ATLAS detector~\cite{PERF-2007-01} is a multi-purpose particle detector
with a forward-backward symmetric cylindrical geometry and a near $4\pi$ coverage in
solid angle.\footnote{ATLAS uses a right-handed coordinate system with its origin at the
nominal interaction point (IP) in the centre of the detector and the $z$-axis along the beam
pipe. The $x$-axis points from the IP to the centre of the LHC ring, and the $y$-axis points
upwards. Cylindrical coordinates $(r,\phi)$ are used in the transverse plane, $\phi$ being the
azimuthal angle around the $z$-axis. The pseudorapidity is defined in terms of the polar angle
$\theta$ as $\eta = -\ln \tan(\theta/2)$. Angular separation is measured in units of
$\Delta R \equiv \sqrt{(\Delta\eta)^2 + (\Delta\phi)^2}$.}
ATLAS comprises an inner detector (ID) surrounded by a thin superconducting solenoid providing 
a \SI{2}{\tesla} axial magnetic field, a calorimeter system and a
muon spectrometer in a toroidal magnetic field. The ID tracking system covers the pseudorapidity
range $|\eta| < 2.5$ and consists of silicon pixel, silicon microstrip, and 
transition radiation tracking detectors. 
The ID provides precise position and momentum measurements for 
charged particles and allows efficient identification of jets containing $b$-hadrons.
Lead/liquid-argon (LAr) sampling calorimeters provide electromagnetic (EM) energy measurements
with high granularity up to $|\eta|=2.5$. A hadron (steel/scintillator-tile) calorimeter covers the central
pseudorapidity range ($|\eta| < 1.7$). The end-cap and forward regions are instrumented with 
LAr calorimeters for both the EM and hadronic energy measurements up to $|\eta| = 4.9$.
The muon spectrometer surrounds the calorimeters. It consists of three large air-core toroid
superconducting magnet systems, separate trigger detectors and high-precision tracking chambers providing 
accurate muon tracking for $|\eta|<2.7$ and muon triggering for $|\eta|<2.4$. 

A three-level trigger system~\cite{PERF-2011-02} is used to select events.
The first-level trigger is implemented in hardware and uses a subset of the detector information 
to reduce the event rate to less than 75~kHz. Two software-based trigger levels, Level-2 and the 
Event Filter, reduce the rate of Level-1 accepts to about 400~Hz on average.

Candidate events are characterised by exactly two leptons
($ee$, $\mu\mu$, $e\mu$), missing transverse 
momentum $\met$ due to the neutrinos from the leptonic decays of the two $W$~bosons, and
a $b$-jet originating from the top-quark decay.
Electron candidates are reconstructed from energy clusters in the calorimeter 
which are matched to ID tracks~\cite{PERF-2013-03}. Selected electrons must have
$\et> 25$~\GeV\,and $|\eta| < 2.47$, excluding the barrel/end-cap transition region of
$1.37<|\eta|<1.52$. A hit in the innermost layer of the ID is required, to reject photon
conversions. Electron candidates are required to fulfil calorimeter-based and track-based
isolation requirements in order to suppress backgrounds from hadron decays.
The calorimeter transverse energy within a cone of size $\Delta R=0.2$ and the scalar sum of
track $\pt$ within $\Delta R$ of $0.3$ around the electron,
in each case excluding the contribution from the electron itself, are each required
to be smaller than $\et$- and $\eta$-dependent thresholds calibrated to give nominal
selection efficiencies of 90\% for prompt electrons from $Z\rightarrow ee$ decays.

Muon candidates are reconstructed by combining matching tracks reconstructed in both the ID
and the muon spectrometer~\cite{PERF-2014-05}. Selected muons have a
$\pt > 25$~\GeV\,and $|\eta|<2.5$. An isolation criterion~\cite{Rehermann:2010vq}
is applied in order to reduce background
contamination from events in which a muon candidate is accompanied by hadrons. 
The ratio of the sum of $\pt$ of additional tracks in a variable-size cone around the
muon, to the $\pt$ of the muon~\cite{Rehermann:2010vq},
is required to be less than 0.05, yielding a selection efficiency of 97\%
for prompt muons from $Z\rightarrow \mu \mu$ decays.

Jets are reconstructed using the anti-$k_{t}$ jet clustering algorithm~\cite{Salam:2007xv} with a 
radius parameter of $R=0.4$, using locally calibrated topological clusters as inputs~\cite{ATLAS-TOPO-CLUSTER}.
Jet energies are calibrated using energy- and $\eta$-dependent correction factors derived from
simulation and with residual corrections from in-situ measurements~\cite{PERF-2011-03}. Jets are required 
to be reconstructed in the range $|\eta| <2.5$ and to have $\pt > 20$~\GeV. To reduce the
contamination due to jets from additional $pp$ interactions in the same or neighbouring bunch crossings
(pileup), tracks originating from the primary vertex must contribute a large fraction to
the scalar sum of the $\pt$ of all tracks in the jet.
This jet vertex fraction (JVF)~\cite{ATLAS-CONF-2014-018} is required to be at least $50 \%$ for jets 
with $\pt<50$~\GeV\,and $|\eta|<2.4$.

To avoid double-counting objects in an event and to suppress leptons from heavy-flavour
decays, overlaps between reconstructed objects are resolved in the following order:
(1) jets overlapping with a selected electron within $\Delta R$ of $0.2$ are
removed; (2) electrons that are within $\Delta R$ of $0.4$ of a jet are removed;
(3) events are rejected if a selected electron shares an ID track
with a selected muon; and (4) muons that are within $\Delta R$ of $0.4$ of a jet are removed. 

The identification of $b$-jets relies of the long
lifetime of $b$-hadrons and the topological properties of secondary and tertiary decay vertices
reconstructed within the jet. A combination of multivariate
algorithms is used to identify $b$-jets ($b$-tag)~\cite{ATLAS-CONF-2011-102}.
The $b$-tag algorithm has an average efficiency of $70\%$ for $b$-jets from $\ttbar$ decays
and an average mis-tag rate of $0.8 \%$~\cite{ATLAS-Perf-btag,ATLAS-CONF-2014-004} for light-quark jets.

The missing transverse momentum ($\etmiss$) is calculated as the magnitude of the vector sum over the energies of all clusters in the calorimeters, and is refined by applying object-level corrections to the contributions arising from identified electrons, muons, and jets~\cite{PERF-2011-07}.

%-------------------------------------------------------------------------------
\section{Data and simulated samples}
\label{sec:DataMC}
%-------------------------------------------------------------------------------

The dataset used for this analysis was collected at $\sqrt{s}=8$~\TeV in 2012 by the ATLAS detector 
at the LHC, and corresponds, after data quality requirements, 
to an integrated luminosity of 20.3~fb$^{-1}$. Events are required to 
have fired either a single-electron or single-muon trigger. The electron and muon 
triggers impose a $\pt$ threshold of 24~\GeV, along with isolation requirements 
on the lepton. To recover efficiency for higher $\pt$ leptons, the isolated lepton 
triggers are complemented by triggers without isolation requirements, but with $\pt$ thresholds of 60~\GeV\,and 36~\GeV\,for electrons and muons respectively.

Samples of signal and background events are simulated using various Monte Carlo (MC) 
generators, as summarised in Table~\ref{tab:mc}. The generators used for the estimation of the 
modelling uncertainties are listed together with the reference simulation for the $Wt$ signal 
and the $\ttbar$ background. In addition, PDFs used by each generator and the perturbative 
order in QCD of the respective calculations 
are provided. All simulation samples are normalised to theoretical cross-section predictions. A top-quark mass of 172.5~\GeV\,is used~\cite{ATLAS:2014wva}.

\begin{table}[!h!tp]
\begin{center}
\caption{Monte Carlo generators used to model the $Wt$ signal and the background processes at $\sqrt{s}=8$~\TeV.
The samples marked with a $\dagger$ are used as alternatives for $Wt$ or $t\bar{t}$ to evaluate modelling uncertainties. DR refers to the diagram-removal scheme and DS to the diagram-subtraction scheme to handle the overlap and interference between $Wt$ and $\ttbar$, as discussed in the text.}
\label{tab:mc}
\begin{tabular}{|l|c|c|c|}
\hline
Process                     & Generator                 & PDF & Normalisation \\ [0.15cm]

\hline
\hline
\multirow{2}{*}{$Wt$} 
& {\scshape Powheg}-{\scshape Box} v1.0       & CT10    &    \\
& + {\scshape Pythia} v6.426, DR               & CTEQ6L1 & \\ [0.15cm]

\multirow{2}{*}{$Wt$ $^{\dagger}$} 
& {\scshape Powheg}-{\scshape Box} v1.0                                           & CT10    & \\
& + {\scshape Pythia} v6.426,  DS                              & CTEQ6L1 & 22.37~pb \\ [0.15cm]
                                         
\multirow{2}{*}{$Wt$ $^{\dagger}$} 
& {\scshape Powheg}-{\scshape Box} v1.0                                           & CT10    & (NLO+NNLL) \\ 
& + {\scshape Herwig} v6.520.2, DR     & CT10    &     \\ [0.15cm] 

\multirow{2}{*}{$Wt$ $^{\dagger}$} 
& {\scshape MC@NLO} v4.06   & CT10           &  \\
& + {\scshape Herwig} v6.520.2, DR & CT10      & \\ [0.15cm]

\hline
\hline
\multirow{2}{*}{$t\bar{t}$}               & 
{\scshape Powheg}-{\scshape Box}  v1.0             & CT10             &   \\  
& + {\scshape Pythia} v6.426 & CTEQ6L1 &   \\ [0.15cm] 

\multirow{2}{*}{$t\bar{t}$ $^{\dagger}$}     & 
{\scshape Powheg}-{\scshape Box}  v1.0             & CT10         & 253~pb\\  
& + {\scshape Herwig} v6.520.2 & CT10 &   (NNLO+NNLL) \\ [0.15cm]

\multirow{2}{*}{$t\bar{t}$ $^{\dagger}$}     & 
{\scshape MC@NLO}  v4.06             & CT10         & \\  
& + {\scshape Herwig} v6.520.2 & CT10 &   \\ [0.15cm]
  
\hline
\multirow{2}{*}{$WW$, $WZ$, $ZZ$}         & 
{\scshape Alpgen} v2.1.4            & CTEQ6L1         &  88~pb   \\  
& + {\scshape Herwig} v6.520.2 & CT10 &  (NLO) \\ [0.15cm]

\multirow{2}{*}{$Z (\rightarrow ee, \mu \mu, \tau \tau)$ + jets}         & 
{\scshape Alpgen} v2.1.4            & CTEQ6L1        & 3450~pb  \\  
& + {\scshape Pythia} v6.426 & CTEQ6L1 &  (NNLO) \\ [0.15cm]
                                   
\hline
\end{tabular}
\end{center}
\end{table}
  
The $Wt$ events are simulated using the NLO generator
{\scshape Powheg}-{\scshape Box}~\cite{Alioli:2011as,Alioli:2010xd}, interfaced to 
{\scshape Pythia}~\cite{PYTHIA} for parton showering with the Perugia 2011C set of tuned parameters~\cite{Skands:2010ak}.
In the {\scshape Powheg}-{\scshape Box} event generator, the CT10~\cite{Lai:2010vv} PDFs 
are used, while the CTEQ6L1~\cite{Nadolsky:2008zw} PDFs are used for {\scshape Pythia}. The generation of
$Wt$ events is performed in the 5FNS. The overlap and
interference between $Wt$ and $\ttbar$ is handled using the diagram-removal scheme (DR),
where all doubly resonant NLO $Wt$ diagrams are removed~\cite{Frixione:2008yi}.
An additional sample, generated with the diagram-subtraction scheme (DS), where the
cross-section contribution from Feynman diagrams containing two top quarks is subtracted,
is used to evaluate the uncertainty associated with the modelling of the overlap between
$Wt$ and $\ttbar$~\cite{Frixione:2008yi}.
Two alternative samples are used to determine theory
modelling uncertainties: one using {\scshape MC@NLO}~\cite{Frixione:2002ik} and the other using
{\scshape Powheg}-{\scshape Box}, both interfaced to {\scshape Herwig}~\cite{Corcella:2000bw},
with {\scshape Jimmy} for underlying-event modelling~\cite{Butterworth:1996zw}.

The dominant and largely irreducible $\ttbar$ background is simulated with {\scshape Powheg}-{\scshape Box}, 
using the CT10 NLO PDF set, with parton showering and hadronisation performed with {\scshape Pythia}.
The $\ttbar$ production cross-section is $\sigma_{\ttbar} = 253^{+13}_{-15}$~pb, 
computed at NNLO in QCD, including resummation of 
NNLL soft gluon terms~\cite{Cacciari:2011hy,Beneke:2011mq,Baernreuther:2012ws,Czakon:2012zr,Czakon:2012pz,Czakon:2013goa,Czakon:2011xx}. 

Smaller backgrounds arise from diboson and $Z$+jets production.
The {\scshape Alpgen} LO generator~\cite{Mangano:2002ea}, interfaced to
{\scshape Herwig}, is used to generate diboson events, with the CTEQ6L1 PDF set.
Diboson events are normalised to the NLO prediction~\cite{Campbell:1999ah}.
The $Z$+jets background is generated with {\scshape Alpgen}, interfaced to {\scshape Pythia},
with the CTEQ6L1 PDF set. 
The diboson estimate also accounts for lower cross-section diboson processes, including $H\to WW$. The $Z$+jets events are normalised to the NNLO
prediction~\cite{Melnikov:2006kv}.

The non-prompt or fake-lepton background arises from non-prompt electrons or muons from the weak decay of mesons events, or from events where one or both leptons are mis-identified. This background contribution includes the $t$-channel and $s$-channel single top-quark production modes. The normalisation and shape of the non-prompt or fake-lepton background is determined directly from data,
using the matrix method~\cite{ATLAS-CONF-2014-058}. 
In addition to events from the signal data sample (labelled as ``tight'' events),
a second (``loose'') set enriched with fake leptons is defined by removing the
lepton isolation requirement. Given the probabilities for real and fake leptons
that already passed the loose selection to also pass the tight selection, the number of tight
events with a fake lepton is determined from a linear system of equations.

Generated events are passed through a simulation~\cite{SOFT-2010-01} of the ATLAS detector 
based on {\scshape Geant4}~\cite{Agostinelli:2002hh} 
and reconstructed using the same procedure as for collision data. The alternative $\ttbar$
samples used to evaluate theory modelling uncertainties are instead processed with
the ATLFAST-II~\cite{SOFT-2010-01} simulation, which employs
a parameterisation of the response of the electromagnetic and hadronic calorimeters, and
{\scshape Geant4} for the other detector components. 
The simulations also include the effect of multiple $pp$ collisions per bunch crossing (pileup).

%-------------------------------------------------------------------------------
\section{Event selection}
\label{sec:sel}
%-------------------------------------------------------------------------------

The dilepton selection requires that each event has a high-quality 
reconstructed primary vertex, which must be formed from at least five tracks with $\pt>0.4$~\GeV. 
Each selected event must contain exactly two isolated opposite-sign leptons ($e$, $\mu$)
that originate from the primary vertex,
at least one of which must be associated with a lepton that triggered the event.
In addition, since the $Wt$ signature contains a high-$\pt$ quark from the top-quark decay, 
events are required to have either one jet or two jets. 

Events from $Z$-boson decays (including $Z\to ee$, $Z\to \mu\mu$, and $Z\to \tau\tau$ with
$\tau\to e$~or~$\mu$) are suppressed through requirements on the invariant mass of the
dilepton system as well as on $\etmiss$  and the pseudorapidity of the leptons+jet(s) system.
Events containing same-flavour leptons ($ee$ or $\mu\mu$) are rejected if the invariant mass of
the lepton pair is between 81~\GeV\,and 101~\GeV. Events are also required to have $\etmiss>40$~\GeV,
with the threshold raised to 70~\GeV\,if the invariant mass of the lepton pair is below
120~\GeV. Events containing one electron and one muon are required to have $\etmiss>20$~\GeV,
with the threshold raised to 50~\GeV\,if the invariant mass of the lepton pair is below 80~\GeV.
Since $Wt$ events are more central than $Z$+jets events, the pseudorapidity of the system
of both leptons and all jets, reconstructed from the vectorial sum of lepton and jet momenta,
is required to be $|\eta^{\textrm{sys}}|<2.5$.

Events are categorised into five regions depending on the jet and $b$-tag multiplicities.
The largest number of expected signal events is in the 1-jet region with one $b$-tagged jet,
while events in the two-jet regions with one or two $b$-tags are dominated by $\ttbar$.
These three regions are included in the cross-section fit. Two additional regions are used to 
validate the modelling of the other backgrounds but are not included in the fit. One-jet and 
two-jet events that have zero $b$-tagged jets compose the 0-tag control regions, which are 
enhanced in the other backgrounds. Observed yields and kinematic distributions in the 0-tag 
control regions are studied while choosing the selection cuts; the three regions included in 
the cross-section fit are not part of this optimisation procedure.

The predicted event yields for signal and backgrounds, and their uncertainties, are summarised in Table~\ref{tab:eventYield}. Uncertainties from different sources are added in quadrature, not taking into account
possible correlations. Many of the sources of systematic uncertainty are common to the $Wt$ signal and $\ttbar$ background processes, and correlated between regions (see Section~\ref{sec:sys}).
The numbers of events observed in data and the total predicted yields are compatible
within the uncertainties.
The $Wt$ signal comprises 21\% of the total expected event yield in the 1-jet 1-tag region. 
The main background originates from the production of top-quark pair events, which
accounts  for almost 80\% of the total event yield in the 1-jet 1-tag region.
For the other regions included in the fit, the expected fraction of signal events is smaller,
9\% in the 2-jet 1-tag region
and 3\% in the 2-jet 2-tag region, which is the most enriched in $\ttbar$.
The other backgrounds are small in the 1-jet 1-tag and 2-jet regions where they account
for 2\% of the total event yield.
The 0-tag control regions are enriched in other backgrounds
(diboson, $Z$+jets and non-prompt or fake lepton),
which contribute $40$--$60 \%$ of the total event yield.

\begin{figure}[htbp]
  \begin{center}
\subfigure[]{\includegraphics[width=0.48\textwidth]{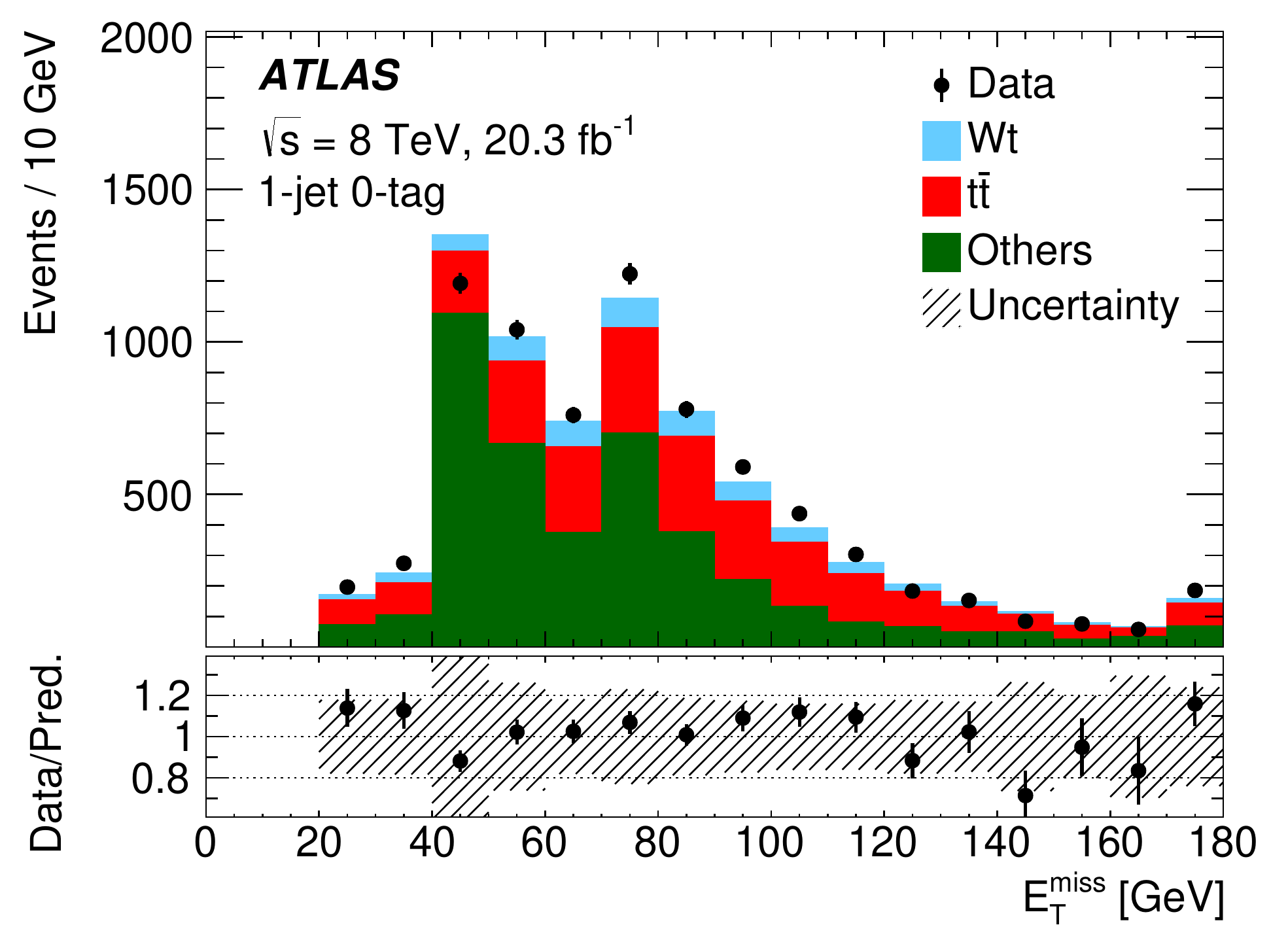}}
\subfigure[]{\includegraphics[width=0.48\textwidth]{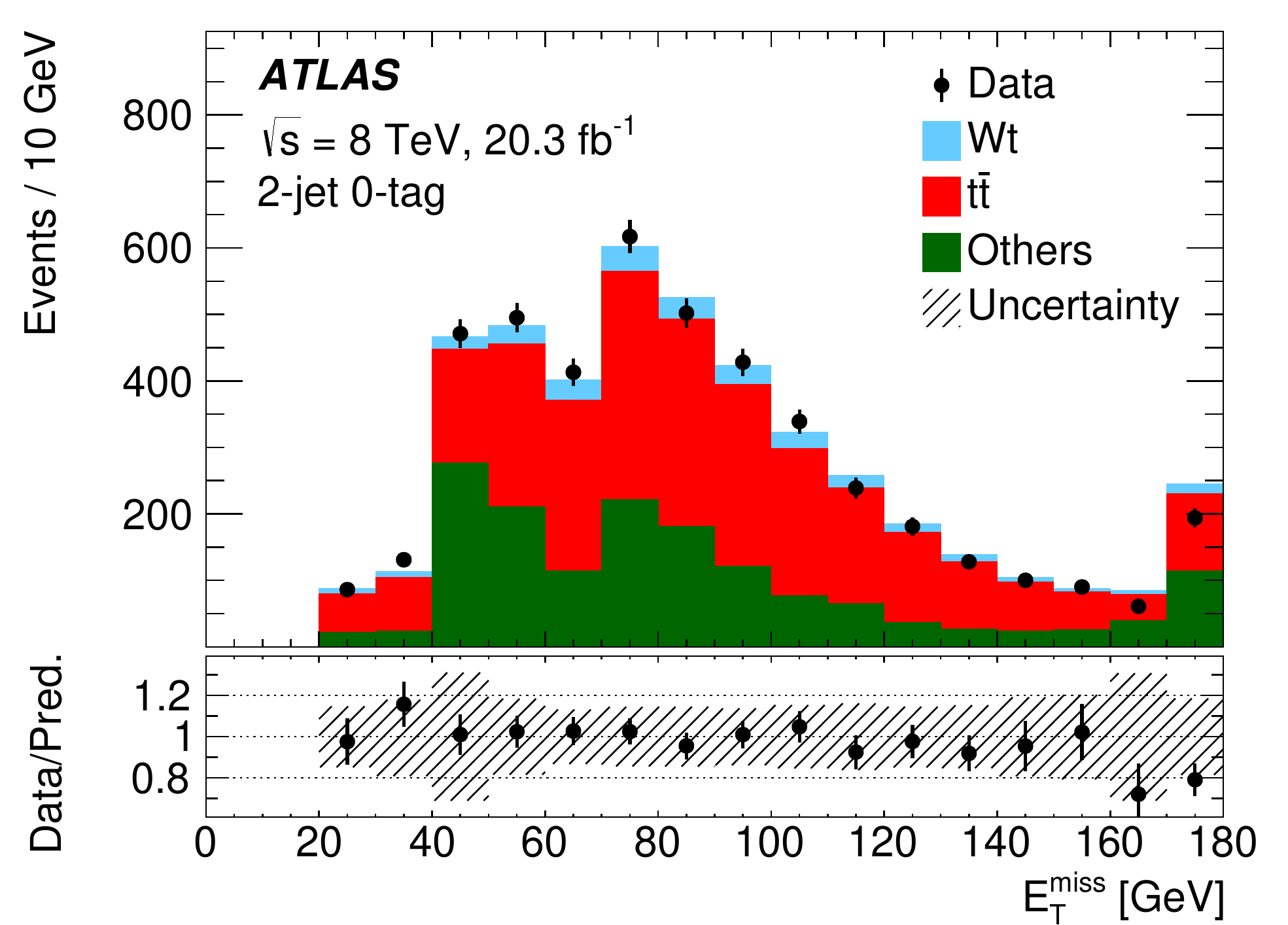}}
\caption{Distributions of the missing transverse momentum $\etmiss$ in
(a) 1-jet and (b) 2-jet events with 0 $b$-tags.
The simulated signal and background contributions are scaled to their expectations.
The hatched area represents the sum in quadrature of the statistical and systematic uncertainties.
The last bin includes the overflow.
\label{fig:CP_0tag}
}
  \end{center}
\end{figure}

\begin{figure}[htbp]
  \begin{center}
\subfigure[]{\includegraphics[width=0.48\textwidth]{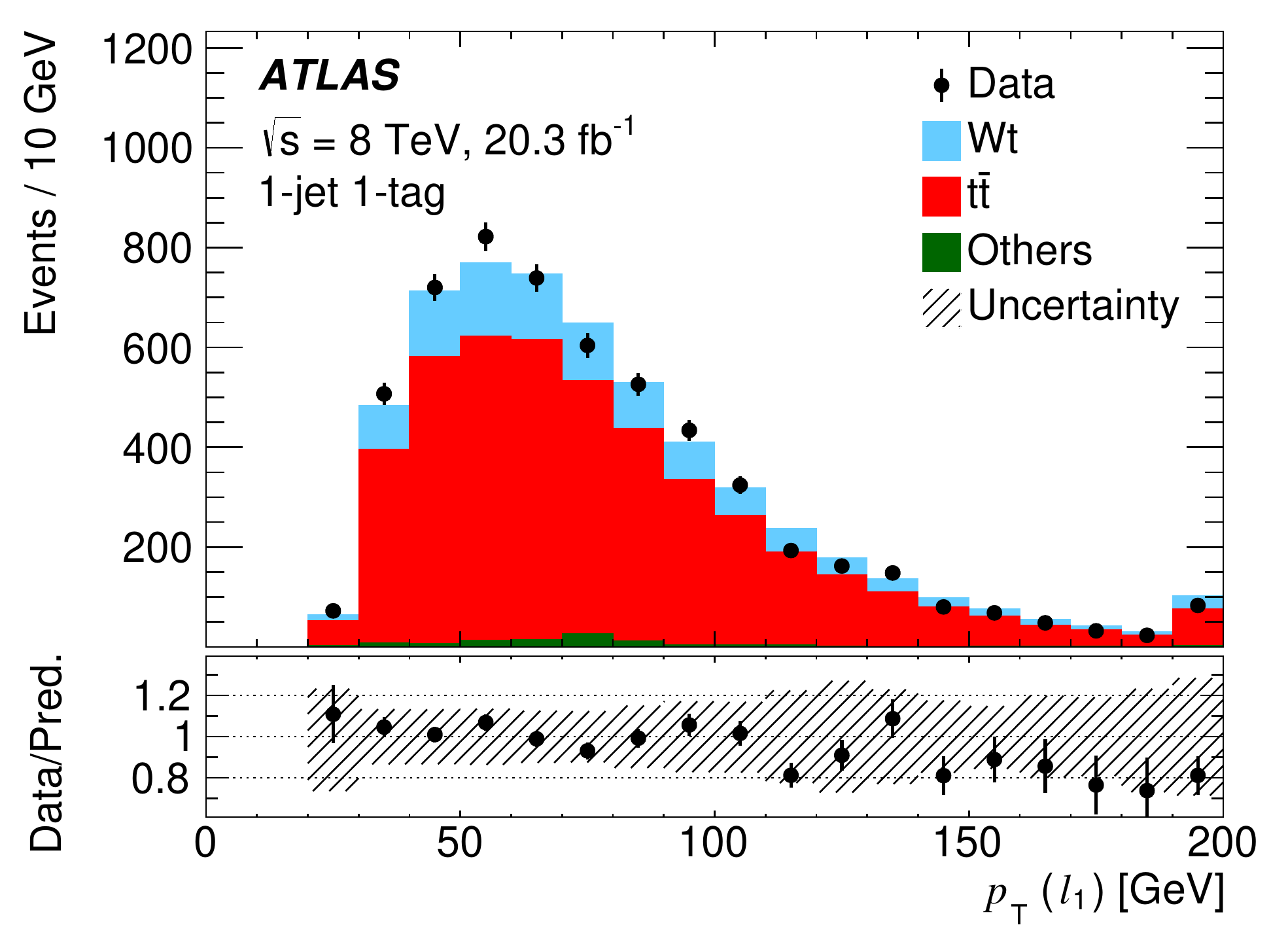}}
\subfigure[]{\includegraphics[width=0.48\textwidth]{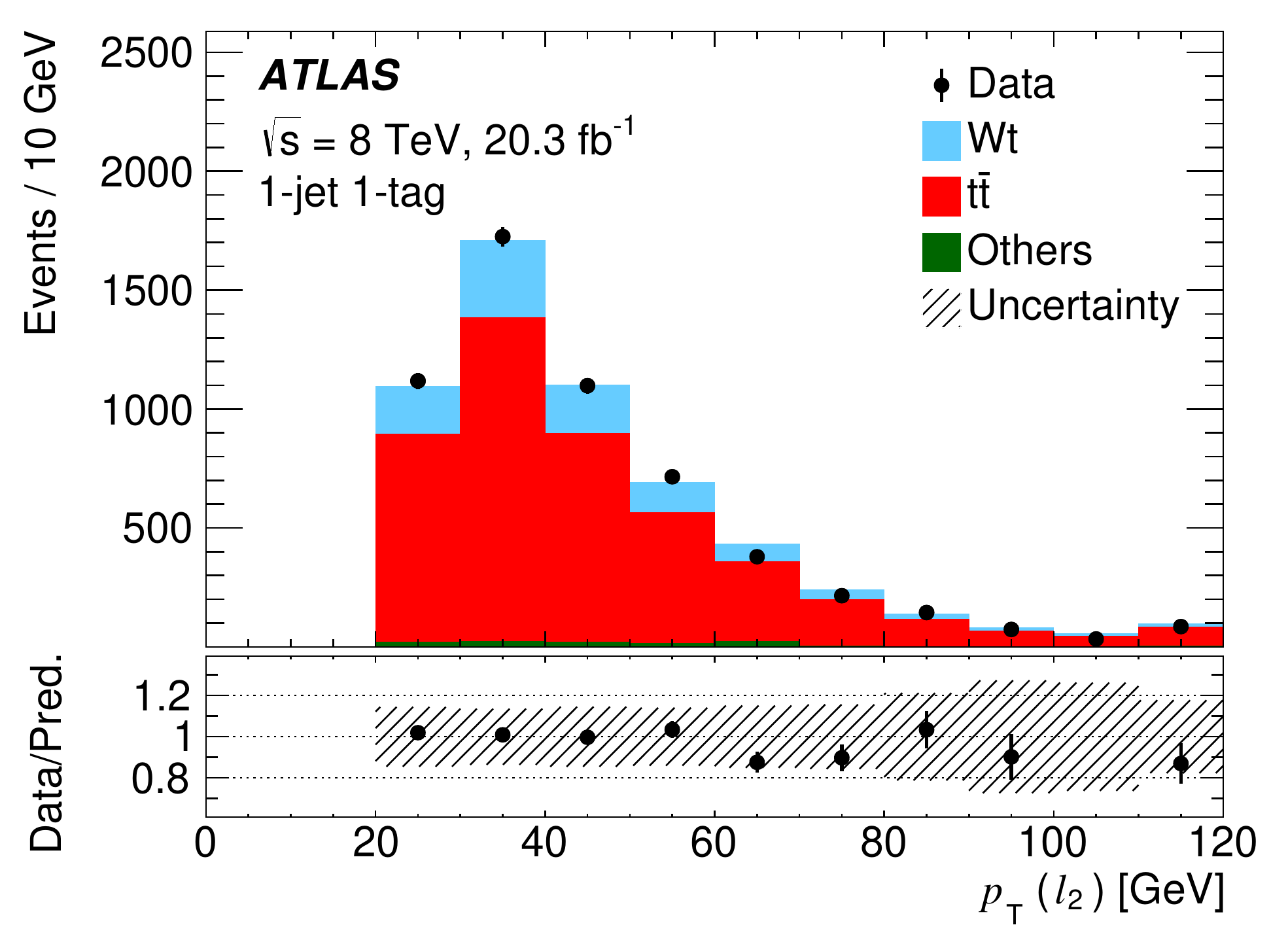}}
\subfigure[]{\includegraphics[width=0.48\textwidth]{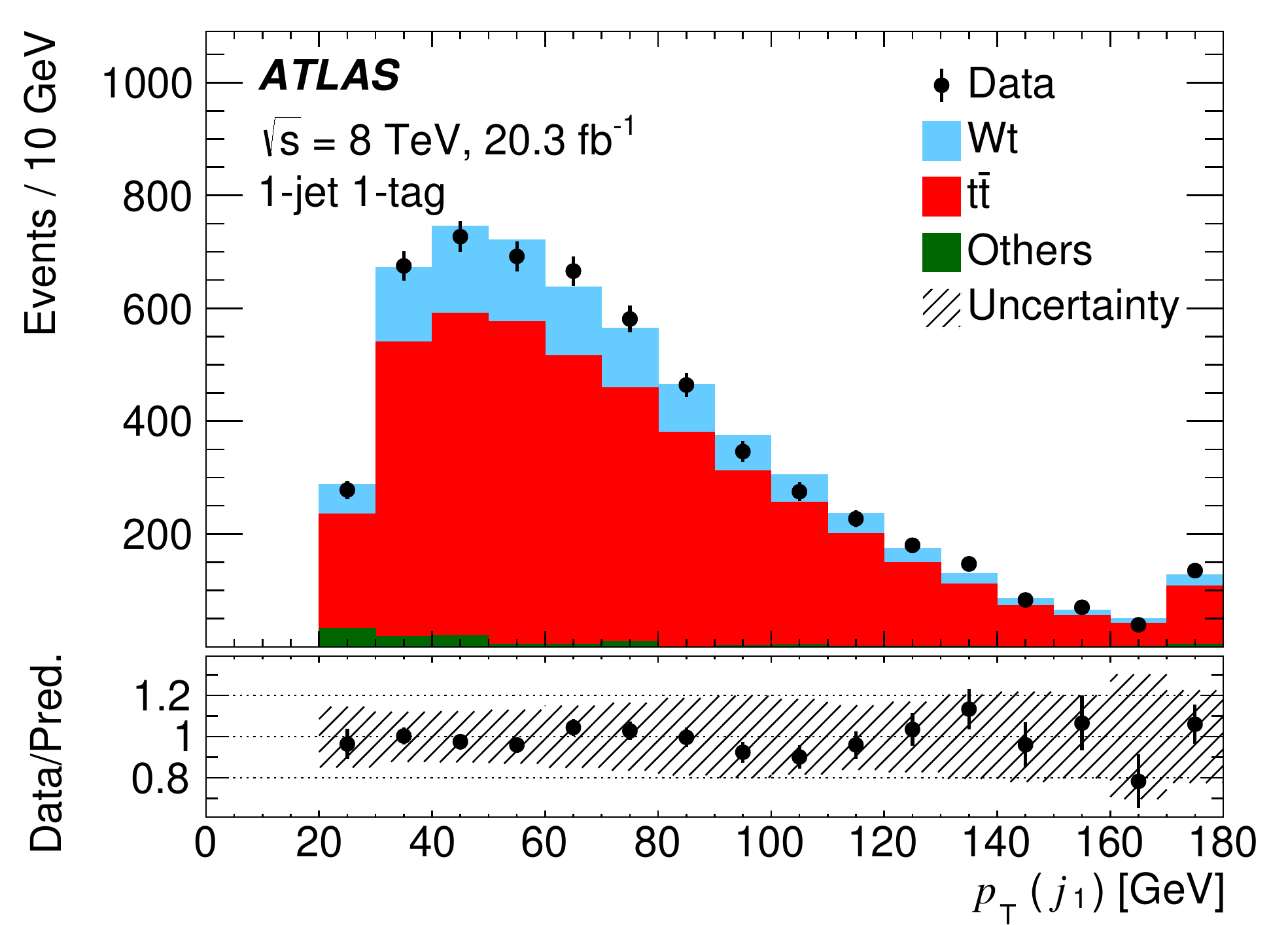}}
\subfigure[]{\includegraphics[width=0.48\textwidth]{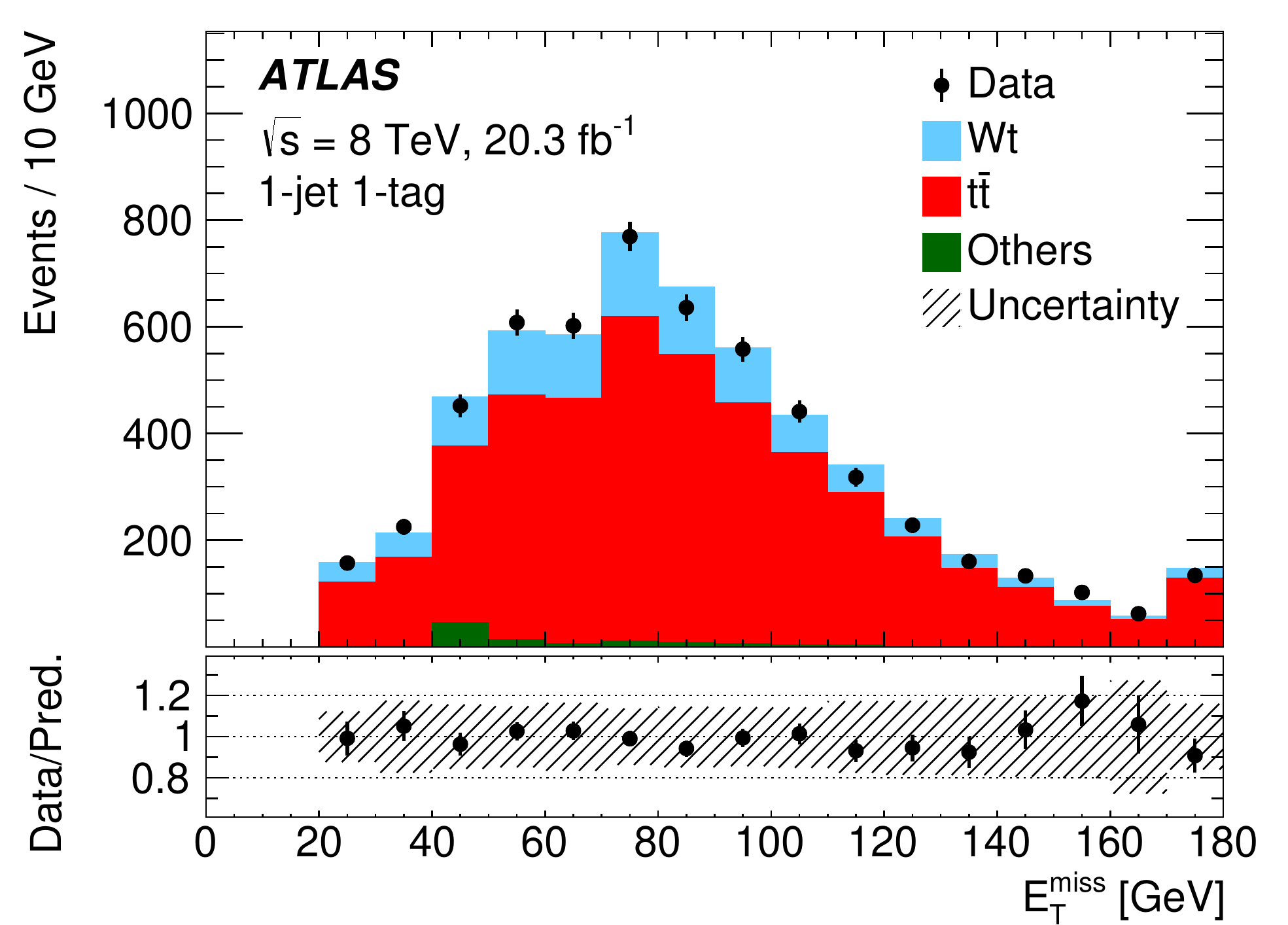}}
\caption{Distributions, in the 1-jet 1-tag region, of
(a) $\pt$ of the leading lepton ($\ell_1$),
(b) $\pt$ of the second-leading lepton ($\ell_2$),
(c) $\pt$ of the jet ($j_1$), and
(d) $\etmiss$.
The simulated signal and background contributions are scaled to their expectations.
The hatched area represents the sum in quadrature of the statistical and
systematic uncertainties. The last bin includes the overflow.
\label{fig:CP_1jet1tag}
}
  \end{center}
\end{figure}

\begin{figure}[htbp]
  \begin{center}
\subfigure[]{\includegraphics[width=0.48\textwidth]{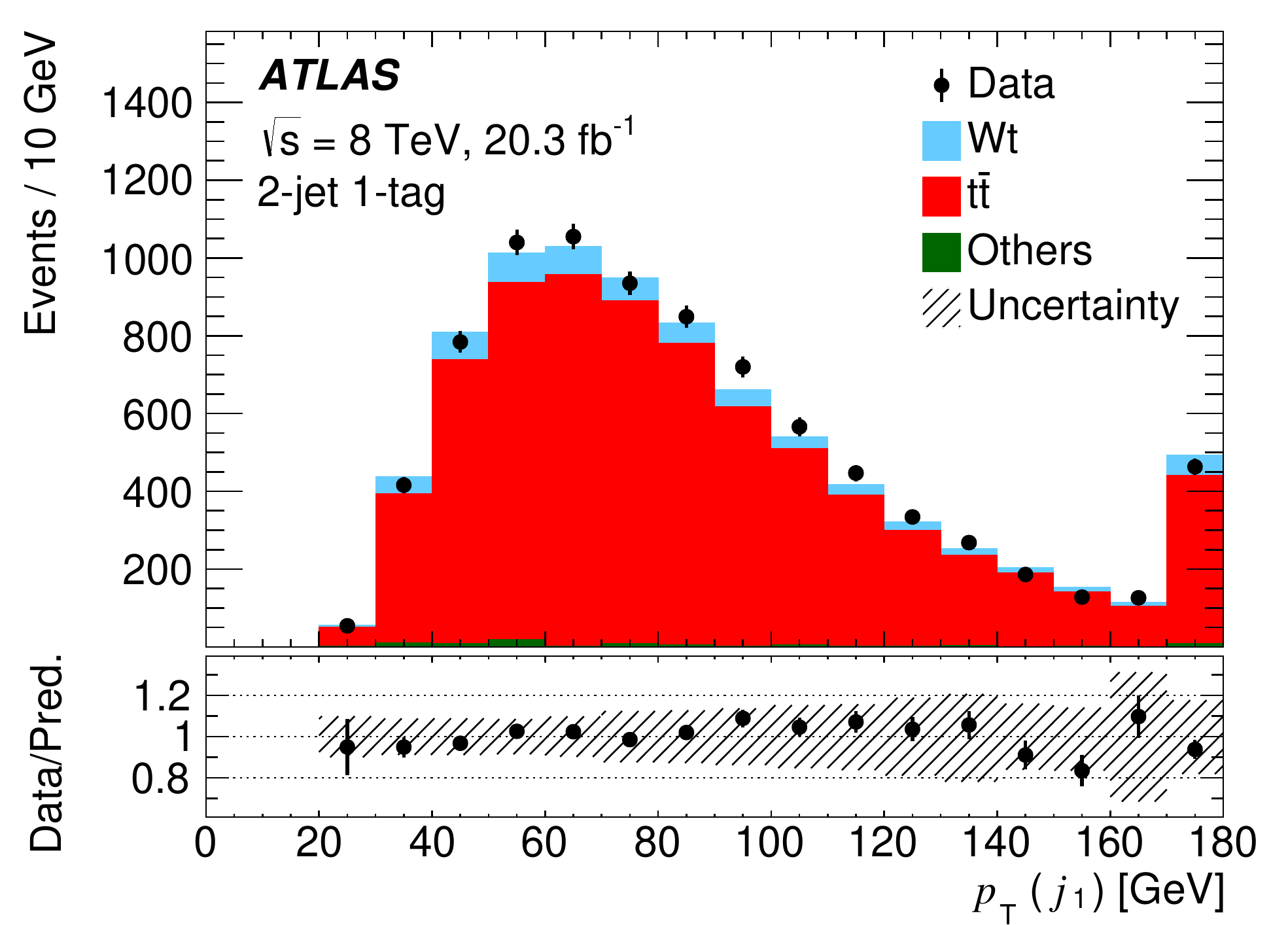}}
\subfigure[]{\includegraphics[width=0.48\textwidth]{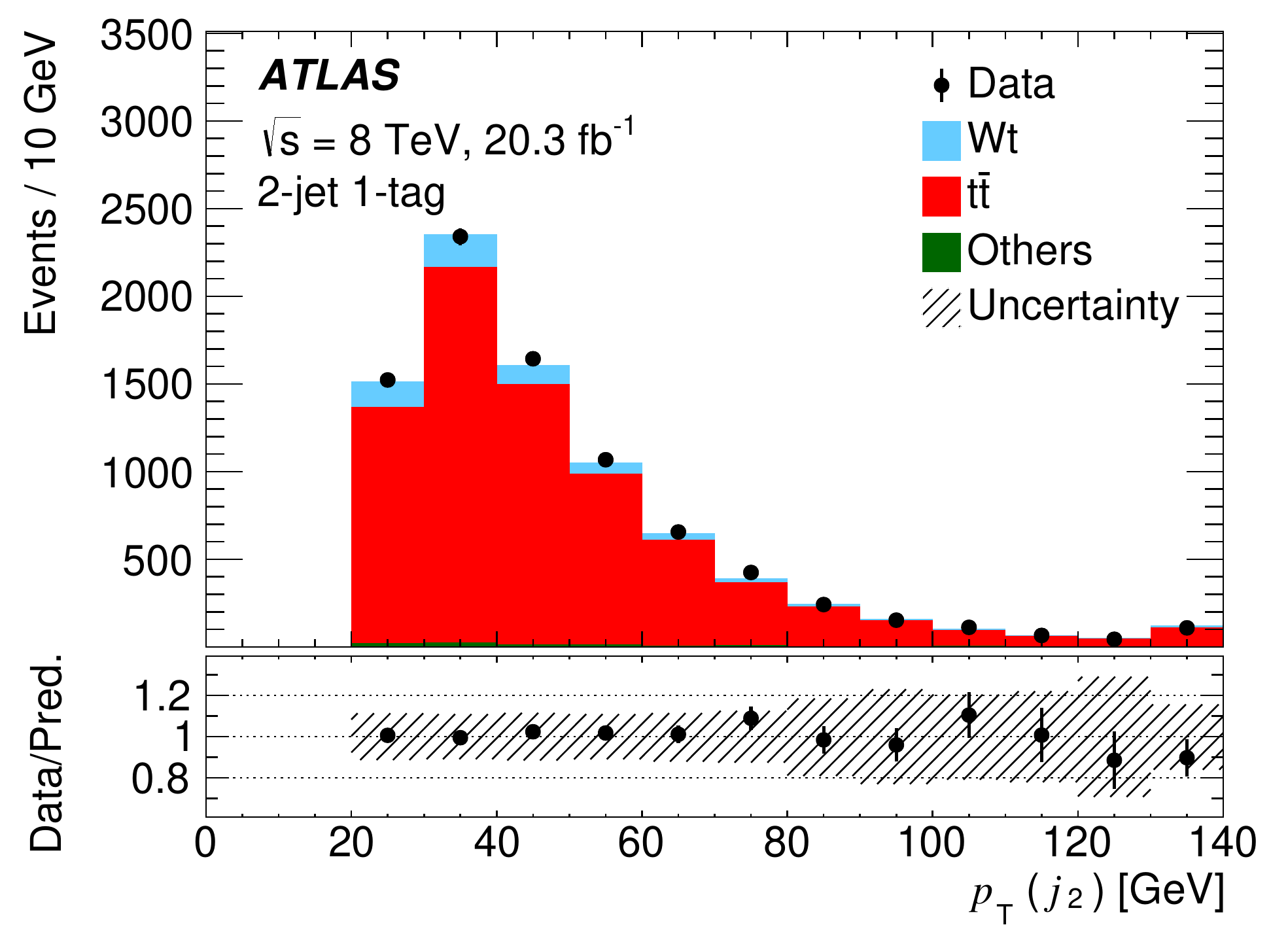}}
\subfigure[]{\includegraphics[width=0.48\textwidth]{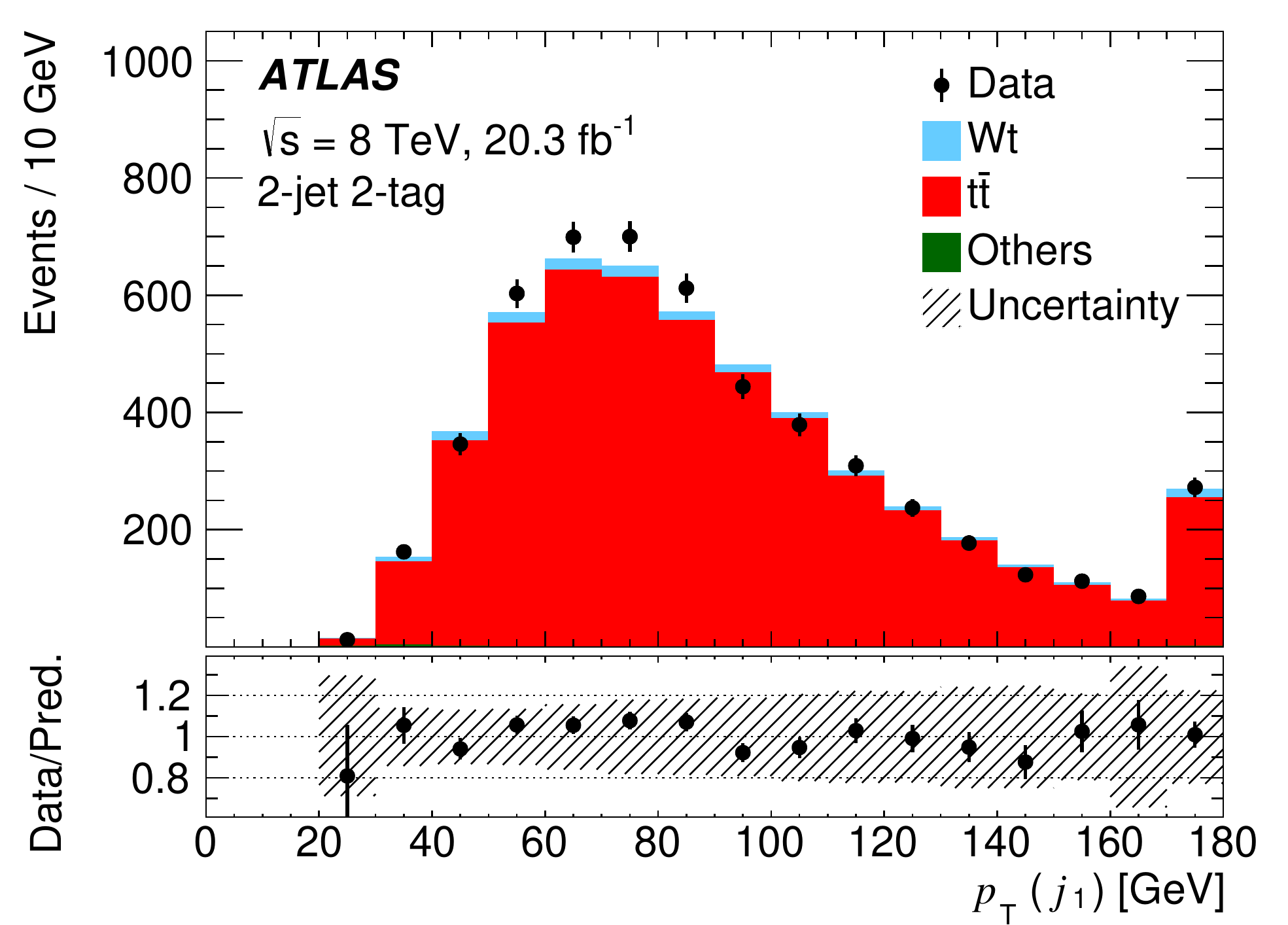}}
\subfigure[]{\includegraphics[width=0.48\textwidth]{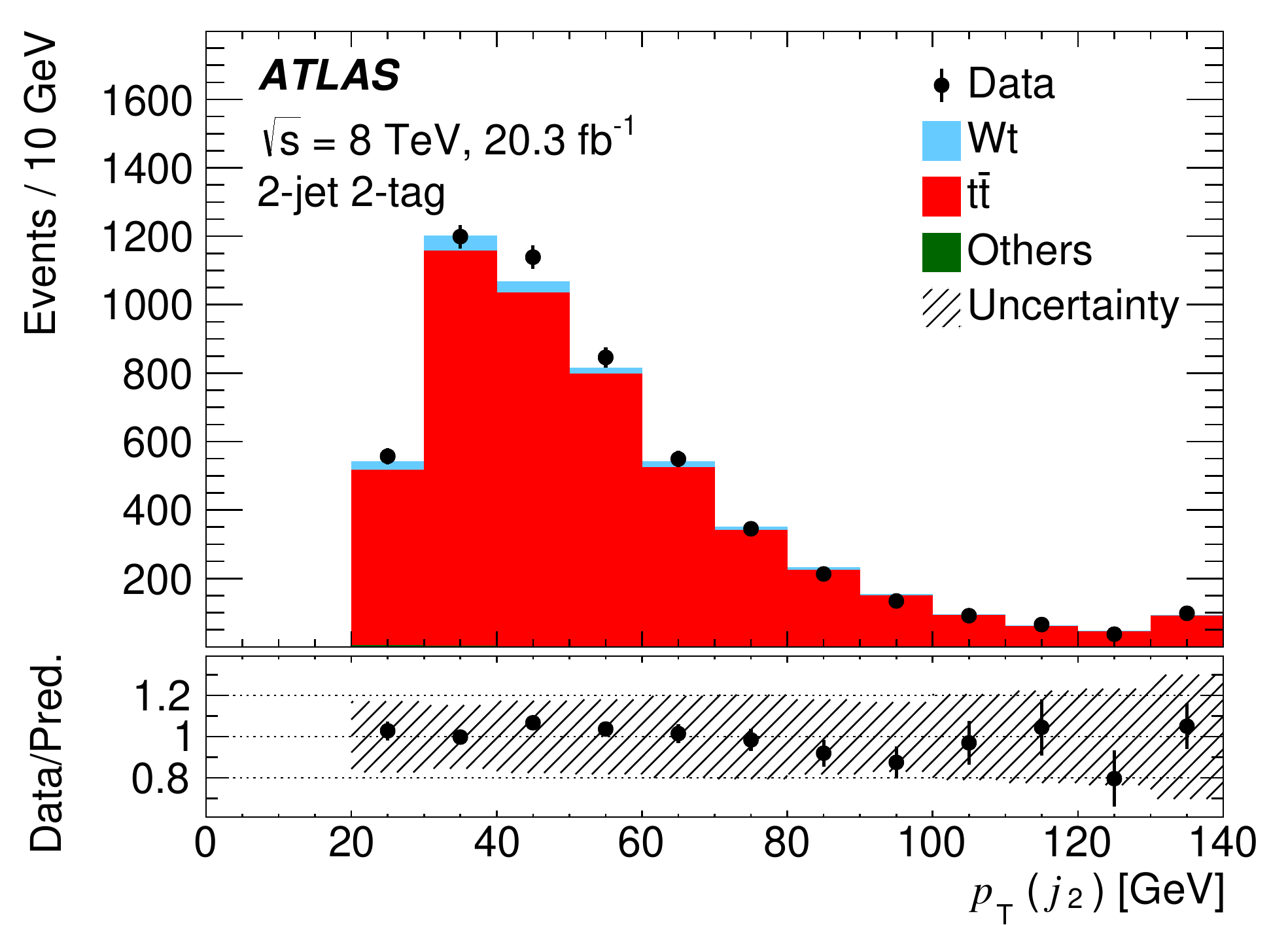}}
\caption{Distributions of the $\pt$ of the leading jet ($j_1$) and the
second-leading jet ($j_2$) in the (a,b) 2-jet 1-tag and (c,d) 2-jet 2-tag regions.
The simulated signal and background contributions are scaled to their expectations.
The hatched area represents the sum in quadrature of the statistical and
systematic uncertainties.
The last bin includes the overflow.
\label{fig:CP_2jet}
}
  \end{center}
\end{figure}

The $\etmiss$ distributions of events in the 0-tag regions are shown in Figure~\ref{fig:CP_0tag} to
demonstrate the good modelling of the other backgrounds. The behaviour of this distribution at
low $\etmiss$ values is a result of the different requirements for same-flavour and
opposite-flavour leptons.
Figures~\ref{fig:CP_1jet1tag} and~\ref{fig:CP_2jet} show the distributions of kinematic
variables of reconstructed objects for the three $b$-tagged regions. 
The data distributions are well modelled by the background and signal 
expectations in all regions.

\begin{table}[!h!tp]
\begin{center}
\caption{Numbers of expected events for the $Wt$ signal and the various background processes
and observed events in data in the five regions, with their predicted uncertainties. Uncertainties shown include all sources of statistical and systematic uncertainty, summed in quadrature.}
\label{tab:eventYield}
\begin{tabular}{|l|
 S[table-figures-uncertainty=1]|
 S[table-figures-uncertainty=1]|
 S[table-figures-uncertainty=1]|
 S[table-figures-uncertainty=1]|
 S[table-figures-uncertainty=1]|
}
  \hline
Process          & {1-jet 1-tag}& {2-jet 1-tag}& {2-jet 2-tag}& {1-jet 0-tag}& {2-jet 0-tag} \\
\hline
\hline

$Wt$             & 1000 \pm 140 &  610 \pm 70 &  160 \pm 50 &  660 \pm 100 &  290 \pm 30  \\ 
$\ttbar$         & 4500 \pm 700 & 7600 \pm 900& 5000 \pm 900& 2600 \pm 400 & 2660 \pm 330 \\ 
Diboson          &   40 \pm 30  &   35 \pm 15 &    1 \pm 1  & 1600 \pm 500 &  670 \pm 270 \\ 
$Z$ + jets       &   70 \pm 40  &   60 \pm 40 &    7 \pm 4  & 2600 \pm 1400&  900 \pm 500 \\
Non-prompt or fake lepton &   24 \pm 15  &   27 \pm 15 &   13 \pm 7  &  130 \pm 70  &   80 \pm 50  \\
\hline
\hline
Total background & 4600 \pm 700 & 7700 \pm 900& 5000 \pm 900& 6900 \pm 1400& 4300 \pm 600 \\
Signal+Background& 5600 \pm 800 & 8300 \pm 900& 5200 \pm 900& 7600 \pm 1500& 4600 \pm 600 \\
\hline
Observed         & 5585         & 8371         & 5273         & 7530         & 4475     \\
  \hline
\end{tabular}
\end{center}
\end{table}

\clearpage
%-------------------------------------------------------------------------------
\section{Analysis}
\label{sec:bdt}
%-------------------------------------------------------------------------------

The separation of the $Wt$ signal from the dominant background from top-quark
pairs is accomplished through the use of a BDT algorithm~\cite{Friedman:2002:SGB:635939.635941}
in the TMVA framework~\cite{Hocker:2007ht}. The BDTs are trained separately in three regions, 1-jet 1-tag, 2-jet 1-tag and 2-jet 2-tag,
using simulated $Wt$ events as signal
and simulated $t\bar{t}$ events as background. Three equal-size $Wt$ samples are combined
to reduce sensitivity to the modelling uncertainties and to maximise the
number of events available for training: 
the {\scshape Powheg-Box+Pythia} sample with the DR scheme, the {\scshape Powheg-Box+Pythia} sample
with the DS scheme, and the {\scshape Powheg-Box+Herwig} sample with the DR scheme. 
The \textsc{AdaBoost} boosting algorithm is used~\cite{adaboost}. 
This algorithm increases the event weight for mis-classified
events for consecutive trees in the training. The final BDT is then the weighted average over all trees.
The list of variables entering the BDT algorithm is chosen based on the power to discriminate
the $Wt$ signal from the $\ttbar$ background and is
derived from a large set of kinematic variables that show good agreement between data and MC simulation. The number of input variables is a compromise between the achievable discrimination power and possible
overtraining. As a result of this optimisation procedure, 13, 16, and 16 variables 
are selected for the 1-jet 1-tag, 2-jet 1-tag, and 2-jet 2-tag regions, respectively.

The BDT input variables used in the three regions are explained below and are
listed in Table~\ref{tab:input-var} together with their importance ranking. 
The objects (denoted $o_{1},\ldots, \, o_n$) used to define these kinematic variables
are the leading- and second-leading lepton ($\ell_1$ and $\ell_2$) and jet ($j_1$ and $j_2$)
as well as $\etmiss$. The kinematic variables are defined as follows. 

\begin{itemize}
\item $\ptsys(o_{1},\ldots, \, o_n)$, magnitude of the vector sum of the transverse momenta of
the objects.

\item \sumet, the scalar sum of transverse energy of calorimeter cells. For cells associated
with electrons and jets, the corresponding corrections are applied.

\item $\sigma$ ($\ptsys(o_{1},\ldots, \, o_n)$), the ratio of $\ptsys$ to
$(H_{\textrm{T}}+\sumet)$, where $H_{\textrm{T}}$ is the scalar sum of the transverse momenta of the
objects.

\item $\Delta \pt(o_1,o_2)$, the difference in $\pt$ between the two objects.

\item $\Delta R(o_1,o_2)$, the separation of the two objects in $\phi$--$\eta$ space.

\item$m_{\textrm{T}}(o_1,o_2)$, the transverse mass, given by
$\sqrt{2\pt(o_1)\pt(o_2)(1-\cos \Delta\phi)}$.

\item Centrality$(o_1,o_2)$, the ratio of the scalar sum of the
$\pt$ of the two objects to the sum of their energies.

\item $m(o_1,o_2)$, the invariant mass of the system of the two objects.

\item $m_{\textrm{T}2}$, which contains information about the presence of the two neutrinos from the two $W$-boson decays~\cite{Lester:1999tx,Barr:2003rg,Cheng:2008hk}. The $m_{\textrm{T}2}$ algorithm creates candidates for the transverse momenta of the two neutrinos, which must sum to give the missing transverse momentum. These are combined with the momenta of the two leptons to
form the transverse mass of two candidate $W$~bosons,
with each also fulfilling a $W$-boson mass constraint. For each such
candidate pair, the larger of the two transverse masses is kept.
Then $m_{\textrm{T}2}$ is given by the smallest transverse mass in all possible candidate pairs.

\item $E/m(o_1,o_2,o_3)$, the ratio of the energy of the system of the three objects to the invariant mass of this system.

\end{itemize}

\begin{table}[!h!tbp]
\begin{center}
\caption{Discriminating variables used in the training of the BDT for each region.
The number indicates the relative importance of this variable, with 1 referring to the
most important variable. An empty field means that this variable is not used in this region.}
\label{tab:input-var}
\begin{tabular}{|l|r|r|r|}
 \hline
 Variable                                                   &  1-jet, 1-tag  & 2-jet 1-tag & 2-jet 2-tag \\ [0.15cm]
 \hline
\hline 

$ \ptsys $ ($\ell_{1},\ell_{2},\etmiss,j_{1}$)               & $ 1$       &             &         \\[0.10cm]
$ \ptsys $ ($\ell_{1},\ell_{2},j_1$)                         & $ 7$       &             &         \\[0.10cm]
$ \ptsys $ ($\ell_{1},\ell_{2}$)                             & $ 13$      &             &         \\[0.10cm]
$ \ptsys$ ($j_{1},j_{2}$)                                &                & $ 10$   & $ 1$ \\[0.10cm]
$ \ptsys$ ($\ell_{1},\ell_{2},\etmiss$)                        &                & $ 12$   & $ 2$ \\[0.10cm]
$ \ptsys$ ($\ell_{1},\ell_{2},\etmiss,j_1,j_2$)                    &                & $ 13$   &          \\[0.10cm]
$ \ptsys$ ($\ell_1,j_1$)                                    &                &             & $ 13$ \\[0.10cm]

$ \sigma (\ptsys) $ ($\ell_{1},\ell_{2},\etmiss,j_{1}$)      & $ 4$       & $ 5$    & \\[0.10cm]

$\pt$ ($j_{2}$)                                            &                &             & $ 8$ \\[0.10cm]
$ \Delta \pt $ ($\ell_{1},\ell_{2}$)                            & $ 8$       &             & \\[0.10cm]
$ \Delta \pt $ (($\ell_1,\ell_2,j_1$),($\etmiss$))              & $ 9$       &             & \\[0.10cm]
$ \Delta \pt$ ($\etmiss,j_{1}$)                           &                & $ 9$    & \\[0.10cm]
$ \Delta \pt$ ($\ell_1,\ell_2$, $\etmiss,j_{1}$)                &                & $ 16$   & \\[0.10cm]
$ \Delta \pt$ ($\ell_2,j_2$)                                   &                &             & $ 14$ \\[0.10cm]

$ \Delta R$ ($\ell_{1},j_{1}$)                                 & $ 2$       &             & $ 5$ \\[0.10cm]
$ \Delta R $ ($\ell_{2}$,$j_{1}$)                              &                & $ 4$    & $ 10$ \\[0.10cm]
$ \Delta R $ ($\ell_{2}$,$j_{2}$)                              &                & $ 6$    & \\[0.10cm]
$ \Delta R $ ($\ell_{2}$,$j_{1}$)                              &                & $ 11$   & \\[0.10cm]
$ \Delta R$ ($\ell_1$,$\ell_2$)                                   &                & $ 14$   & \\[0.10cm]
$ \Delta R $ (($\ell_1,\ell_2$),$j_2$)                             &                &             & $ 9$ \\[0.10cm]

$m$ ($\ell_{2},j_{1}$)                                         & $ 10$      & $ 3$    & $ 3$ \\[0.10cm]
$m$ ($\ell_{1},j_{2}$)                                         &                & $ 1$    & $ 4$ \\[0.10cm]
$m$ ($j_{1},j_{2}$)                                         &                & $ 2$    & \\[0.10cm]
$m$ ($\ell_{2},j_{2}$)                                         &                & $ 7$    & $ 7$ \\[0.10cm]
$m$ ($\ell_{1},j_{1}$)                                         &                & $ 8$    & $ 6$ \\[0.10cm]
$m$ ($\ell_{1},\ell_{2}$)                                         &                & $ 15$   & \\[0.10cm]
$m$ ($\ell_{2},j_{1},j_{2}$)                                   &                &             & $ 11$ \\[0.10cm]
$m$ ($\ell_{1},\ell_{2},j_{1},j_{2}$)                             &                &             & $ 15$ \\[0.10cm]

$ m_{\textrm{T}}$ ($j_{1},\etmiss$)                             & $ 5$       &        & \\[0.10cm]
$ m_{\textrm{T}2}$                                                   & $ 11$      &        & \\[0.10cm]
$E/m$ ($\ell_1,\ell_2,j_2$)                                       &                &        & $ 16$ \\[0.10cm]

$ \sum {E_{\textrm{T}}} $                                            & $ 3$       &             & \\[0.10cm]

Centrality($\ell_{1},\ell_{2}$)                              & $ 6$       &             & \\[0.10cm]
Centrality($\ell_{1},j_1$)                                & $ 12$      &             & \\[0.10cm]
Centrality($\ell_{2},j_{2}$)                              &                &             & $ 12$ \\[0.10cm]

 \hline
\end{tabular}
\end{center}
\end{table}

\begin{figure}[htbp]
  \begin{center}
\subfigure[]{\includegraphics[width=0.48\textwidth]{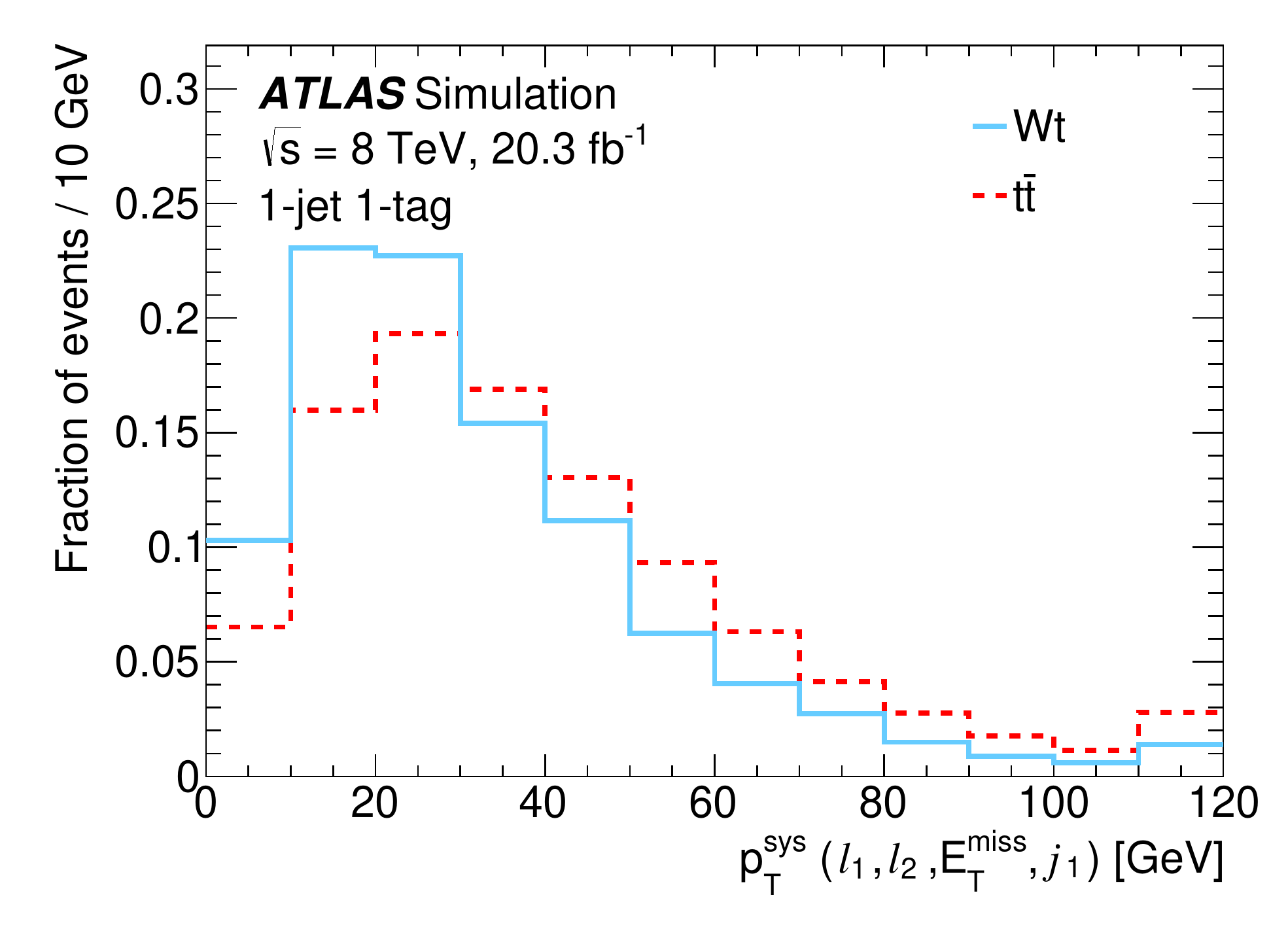}}
\subfigure[]{\includegraphics[width=0.48\textwidth]{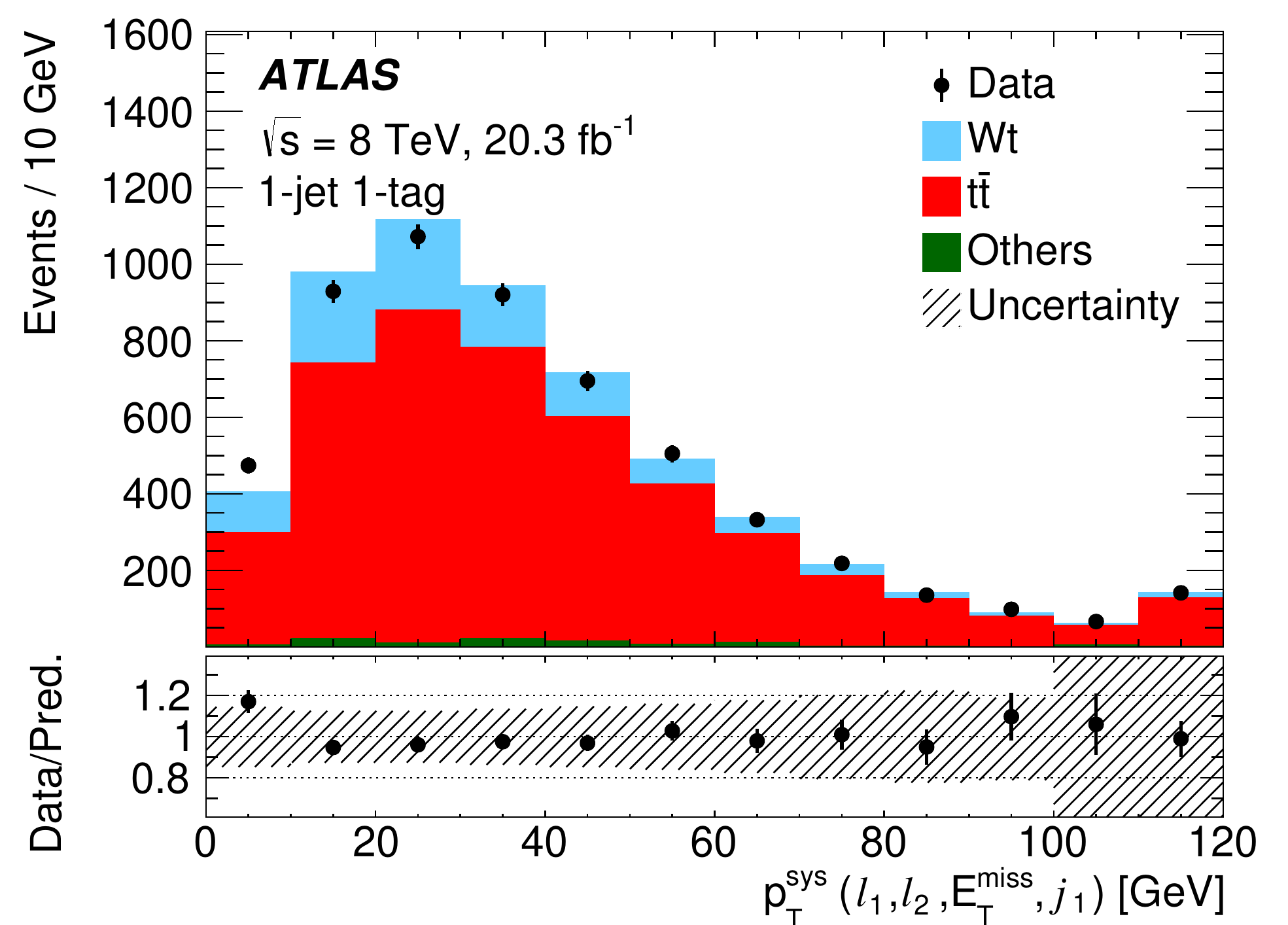}}
\subfigure[]{\includegraphics[width=0.48\textwidth]{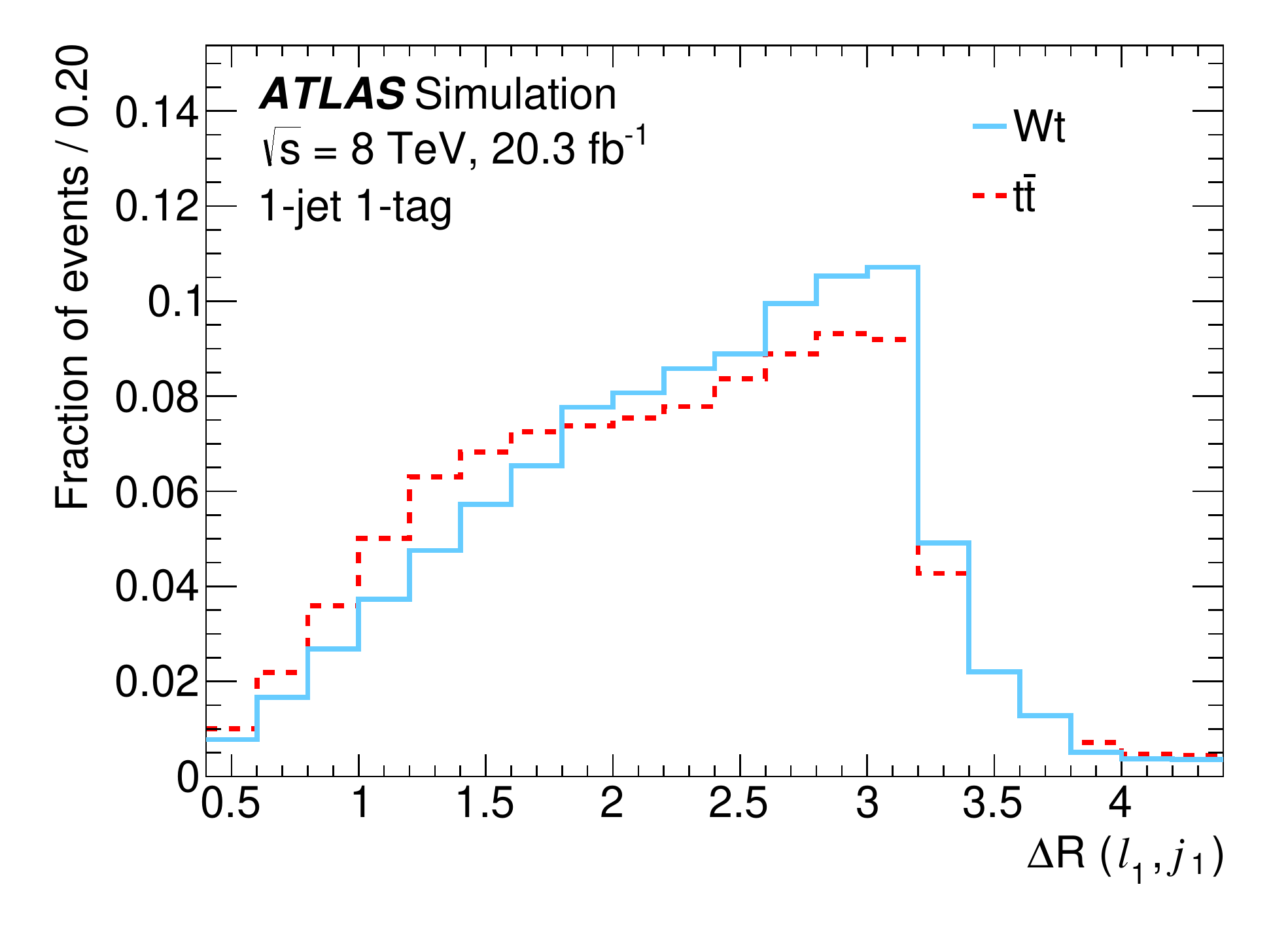}}
\subfigure[]{\includegraphics[width=0.48\textwidth]{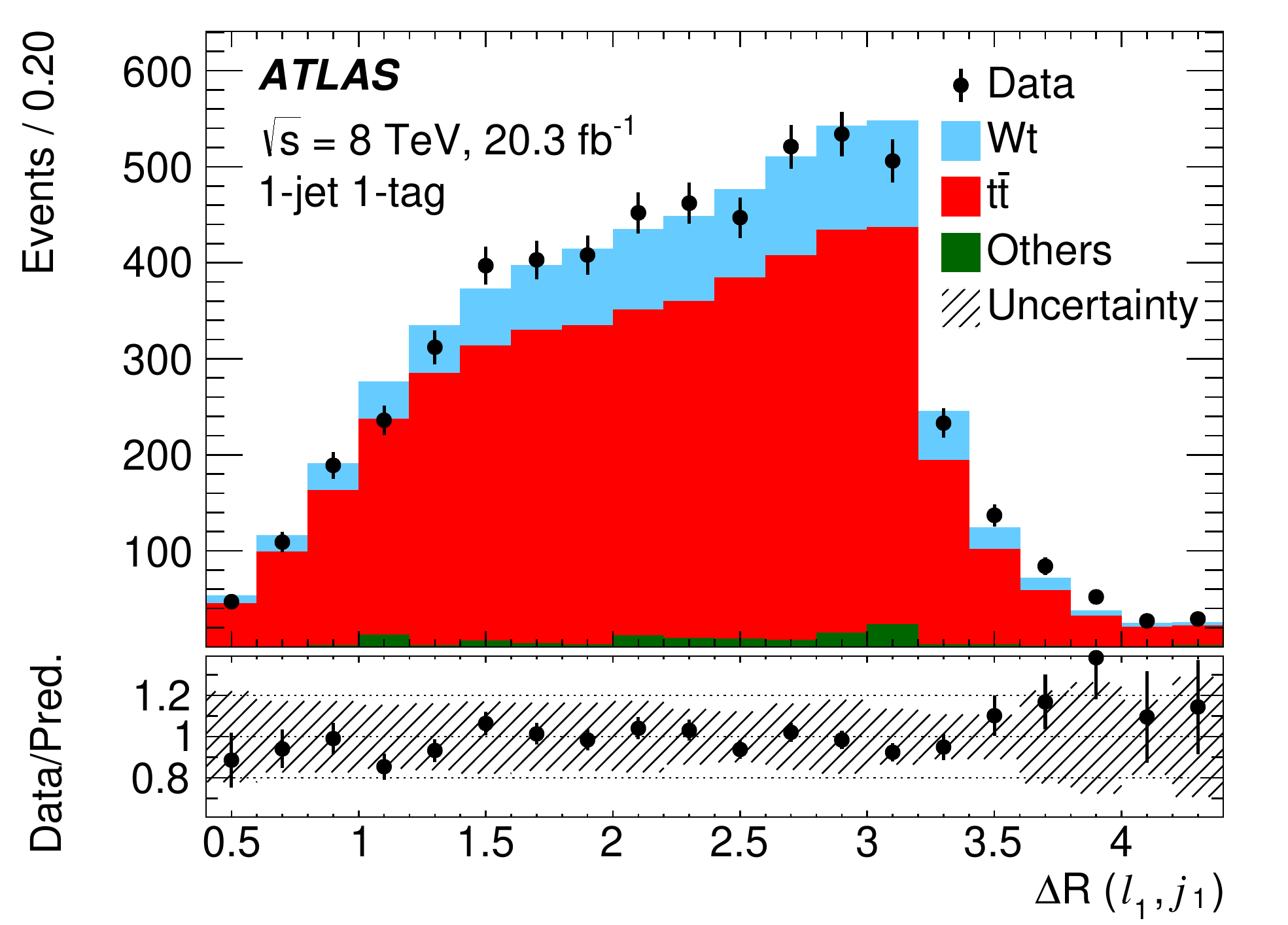}}
\caption{Distributions of the two most important BDT input variables for the 1-jet 1-tag region. 
The distributions are shown for (a, b) the $\pt$ of the system of the leptons, jet and
$\etmiss$ and (c, d) the $\Delta R$ between the leading lepton and the jet.
Each contribution is normalised to unit area in (a, c) and to its expectation in (b, d).
The hatched area represents the sum in quadrature of the statistical and
systematic uncertainties.
The last bin includes the overflow.
\label{fig:IV_1jet}
}
  \end{center}
\end{figure}

\begin{figure}[htbp]
  \begin{center}
\subfigure[]{\includegraphics[width=0.48\textwidth]{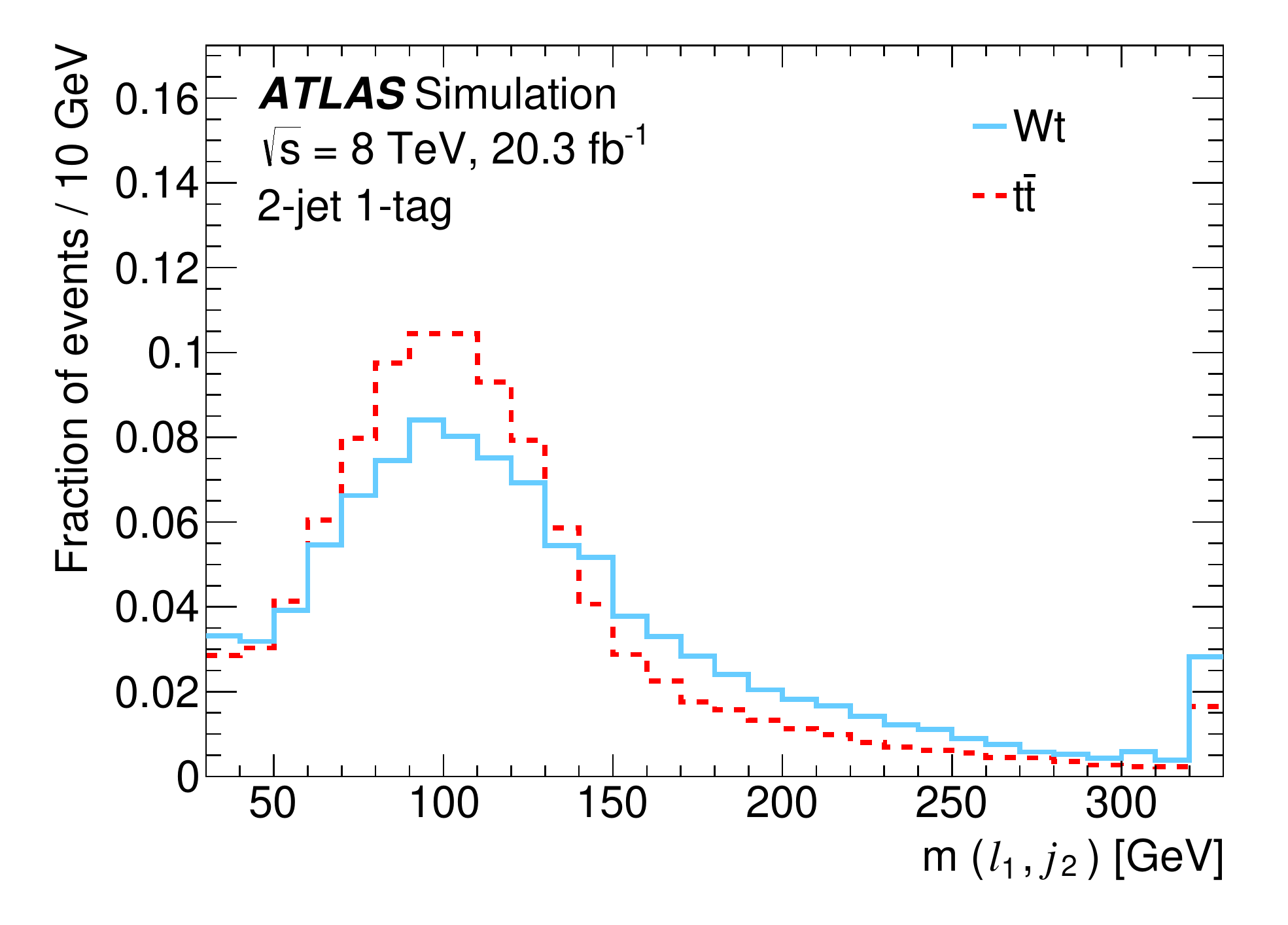}}
\subfigure[]{\includegraphics[width=0.48\textwidth]{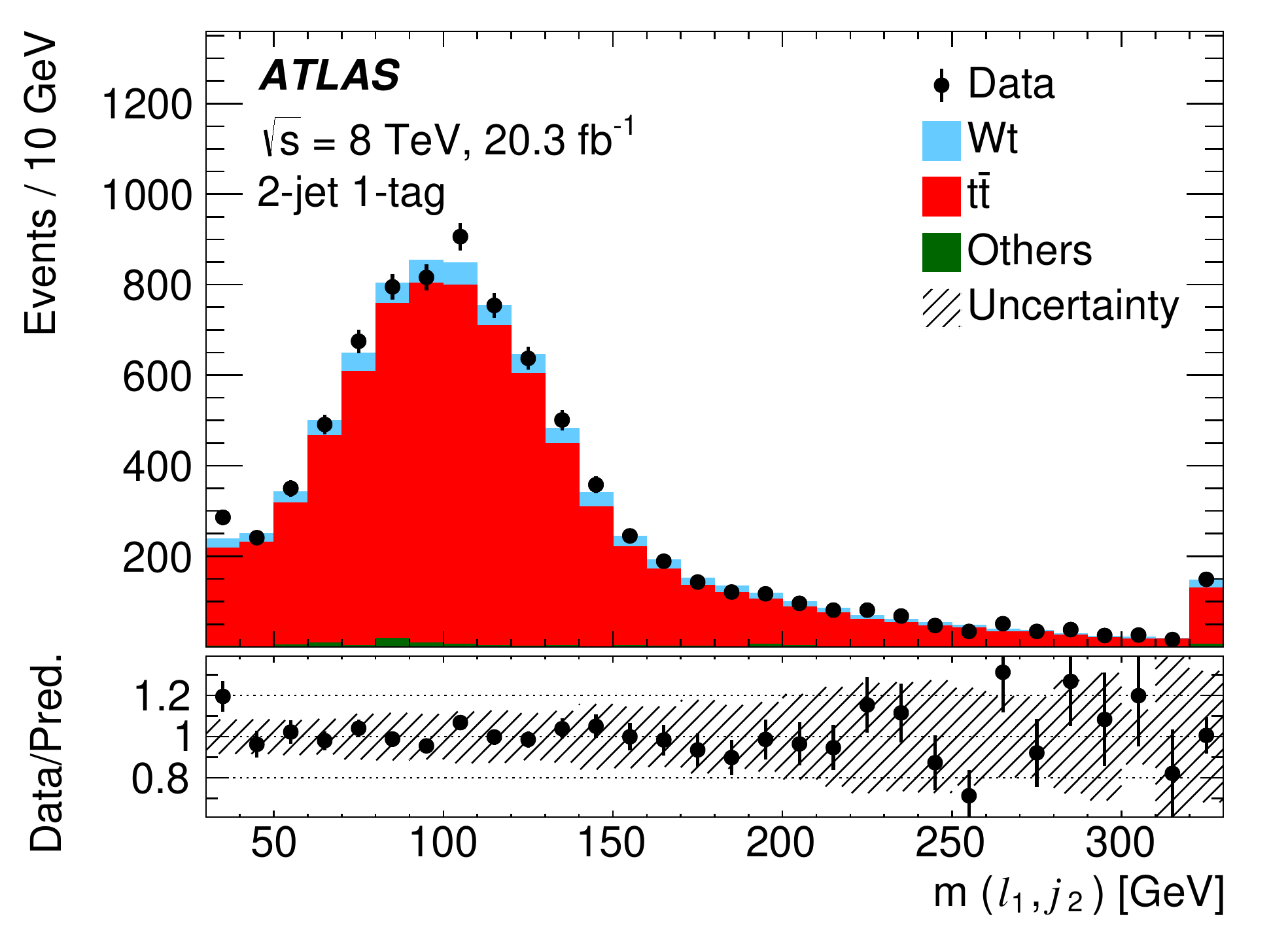}}
\subfigure[]{\includegraphics[width=0.48\textwidth]{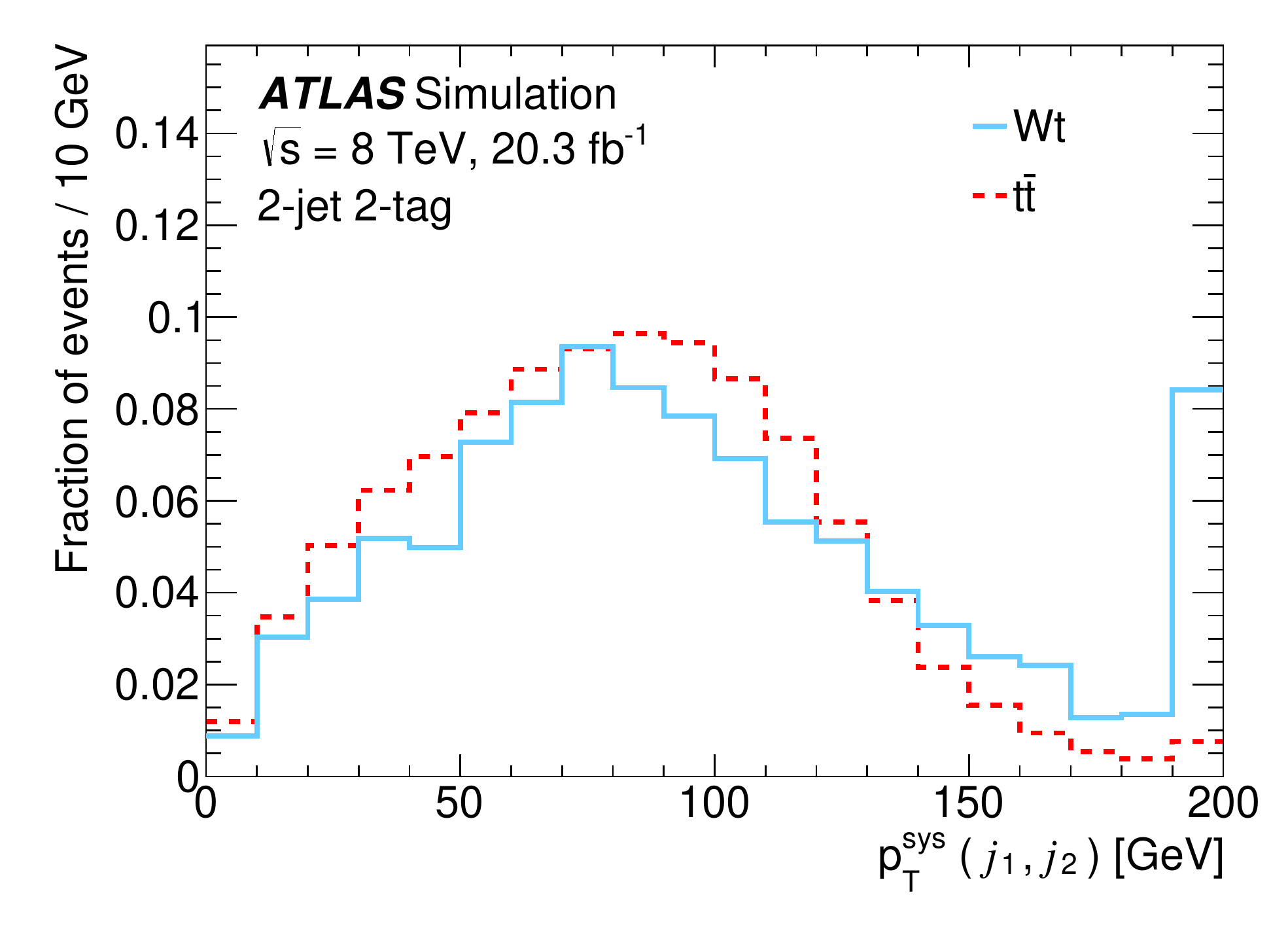}}
\subfigure[]{\includegraphics[width=0.48\textwidth]{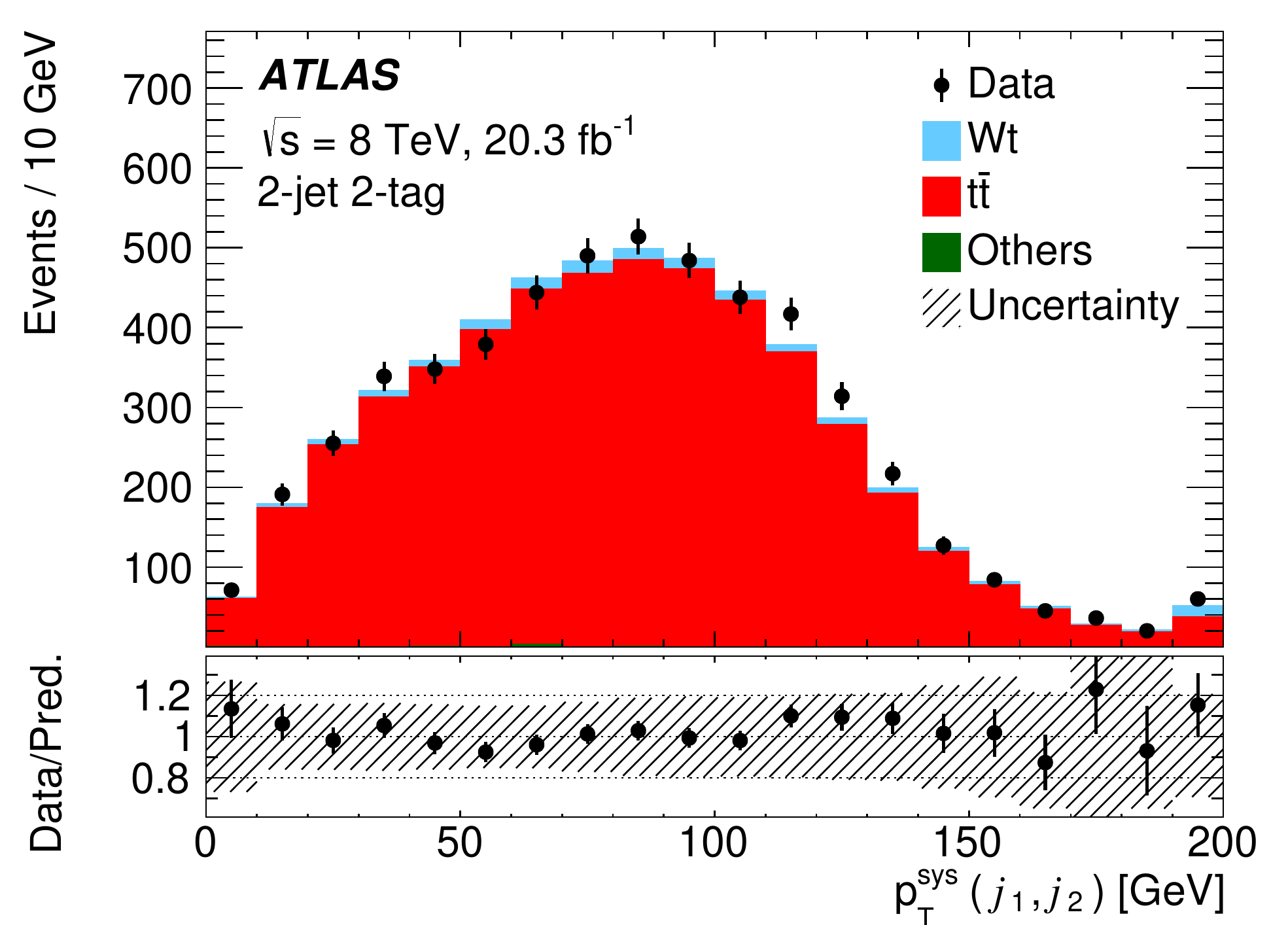}}
\caption{Distributions of the most important BDT input variables in the (a, b) 2-jet 1-tag 
and (c, d) 2-jet 2-tag regions.
The distributions are shown for (a, b) the invariant mass of the system of the leading lepton and the second-leading jet and (c, d) the $\pt$ of the system of the two jets.
Each contribution is normalised to unit area in (a, c) and to its expectation in (b, d).
The hatched area represents the sum in quadrature of the statistical and
systematic uncertainties.
The last bin includes the overflow.
\label{fig:IV_2jet}
}
  \end{center}
\end{figure}

Figure~\ref{fig:IV_1jet} compares the shapes of the most important variables in
the 1-jet 1-tag region for $Wt$ and $\ttbar$ events and shows a comparison of the data and the
SM predictions. The most important variable is
$\ptsys$($\ell_{1},\ell_{2},\etmiss,j_{1}$), which is sensitive to the unidentified $b$-quark
in $t\bar{t}$ events. This variable peaks at lower values for $Wt$ and has a
longer tail for $\ttbar$. The second most important variable is the separation of the
leading lepton and the jet, in $\phi$-$\eta$ space. These two objects originate from the same
top quark in $Wt$ events, leading to a sharper peak than in $\ttbar$ events.
Figure~\ref{fig:IV_2jet} shows the most important discriminating variables in the 2-jet regions.
Here, the $\ptsys$ distribution also peaks at lower values for $Wt$ than for
$\ttbar$, but the distribution is also broader for $Wt$, resulting in a long tail.
The invariant mass variables are important for 2-jet events, where half of the possible
lepton--jet pairings correspond to the objects from the decay of one of the top quarks in
$\ttbar$ events leading to a peak at lower invariant mass.
For $Wt$, only one quarter of the
possible pairings of jets and leptons correspond to the objects from the top-quark decay.

\begin{figure}[htbp]
  \begin{center}
\subfigure[]{
\includegraphics[width=0.48\textwidth]{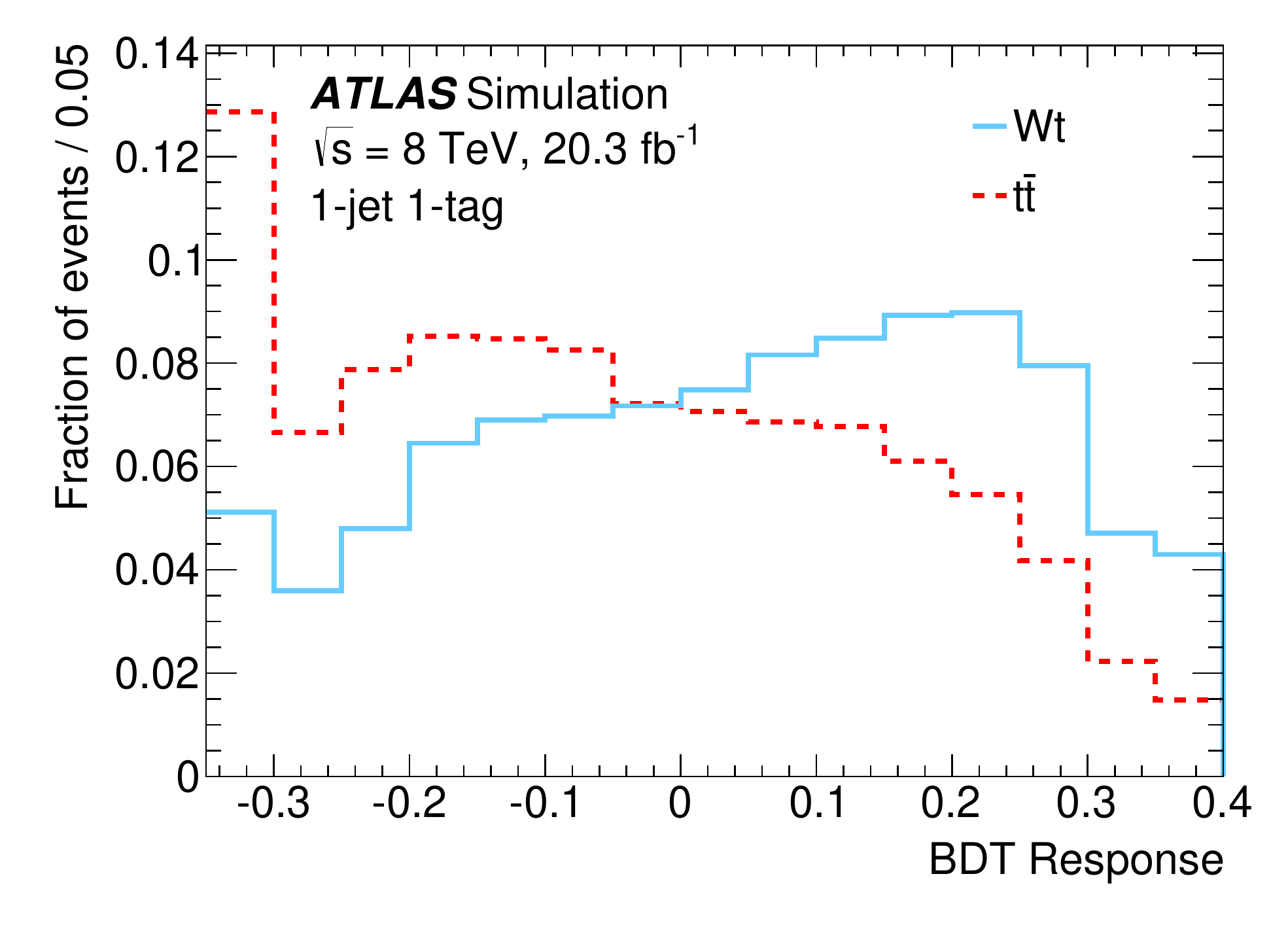}
}
\subfigure[]{\includegraphics[width=0.48\textwidth]{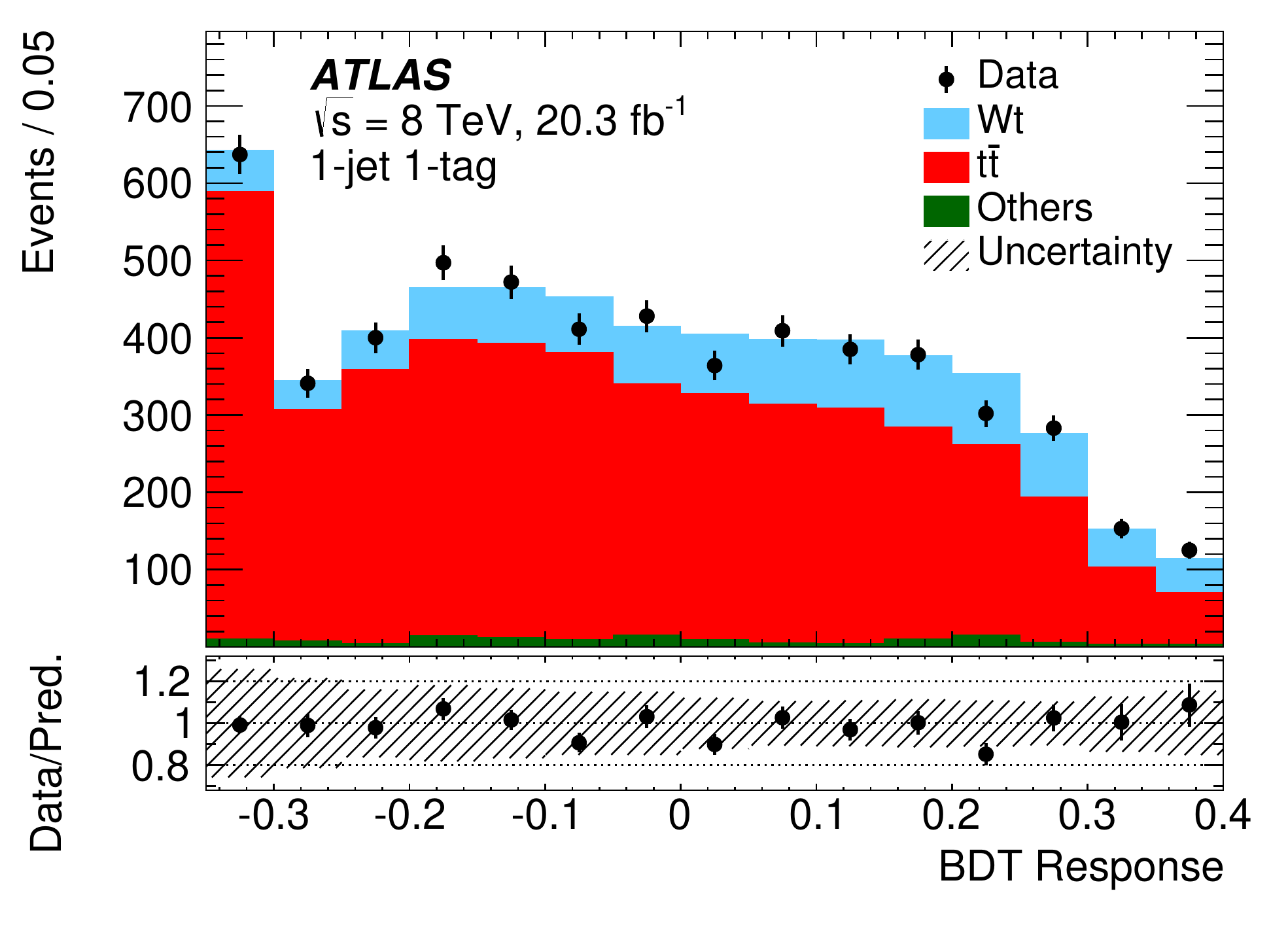}}
\subfigure[]{
\includegraphics[width=0.48\textwidth]{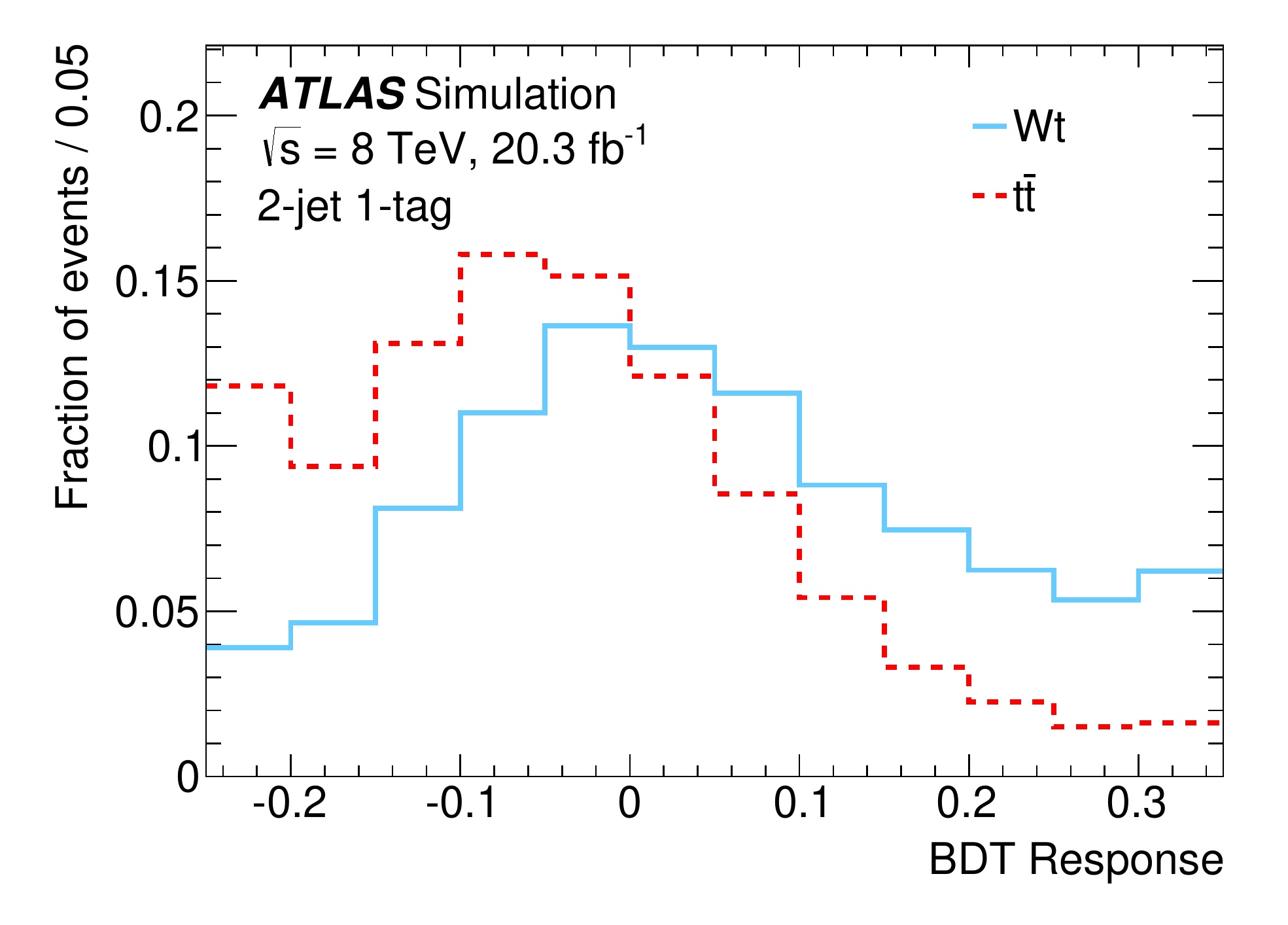}
}
\subfigure[]{\includegraphics[width=0.48\textwidth]{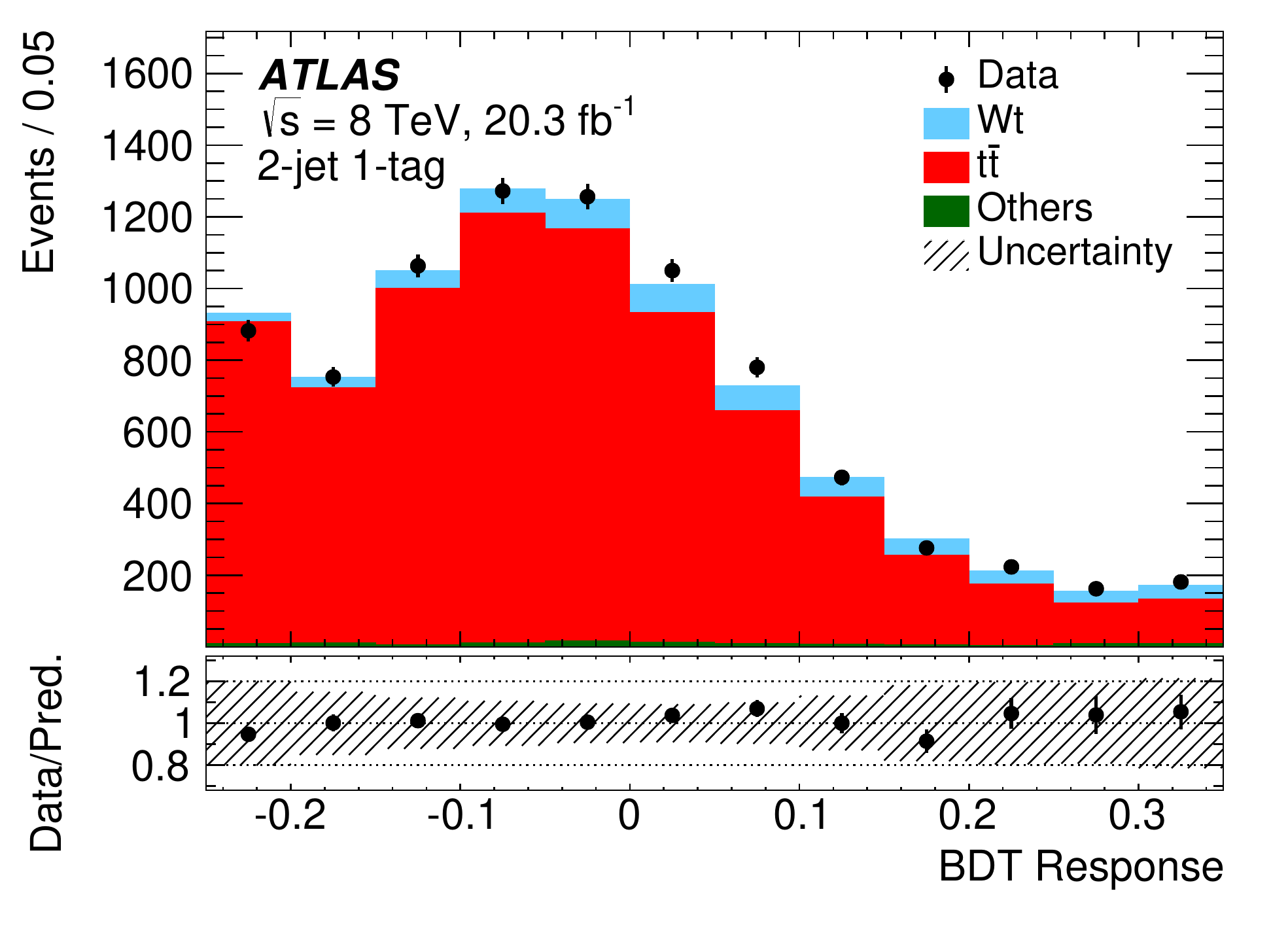}}
\subfigure[]{
\includegraphics[width=0.48\textwidth]{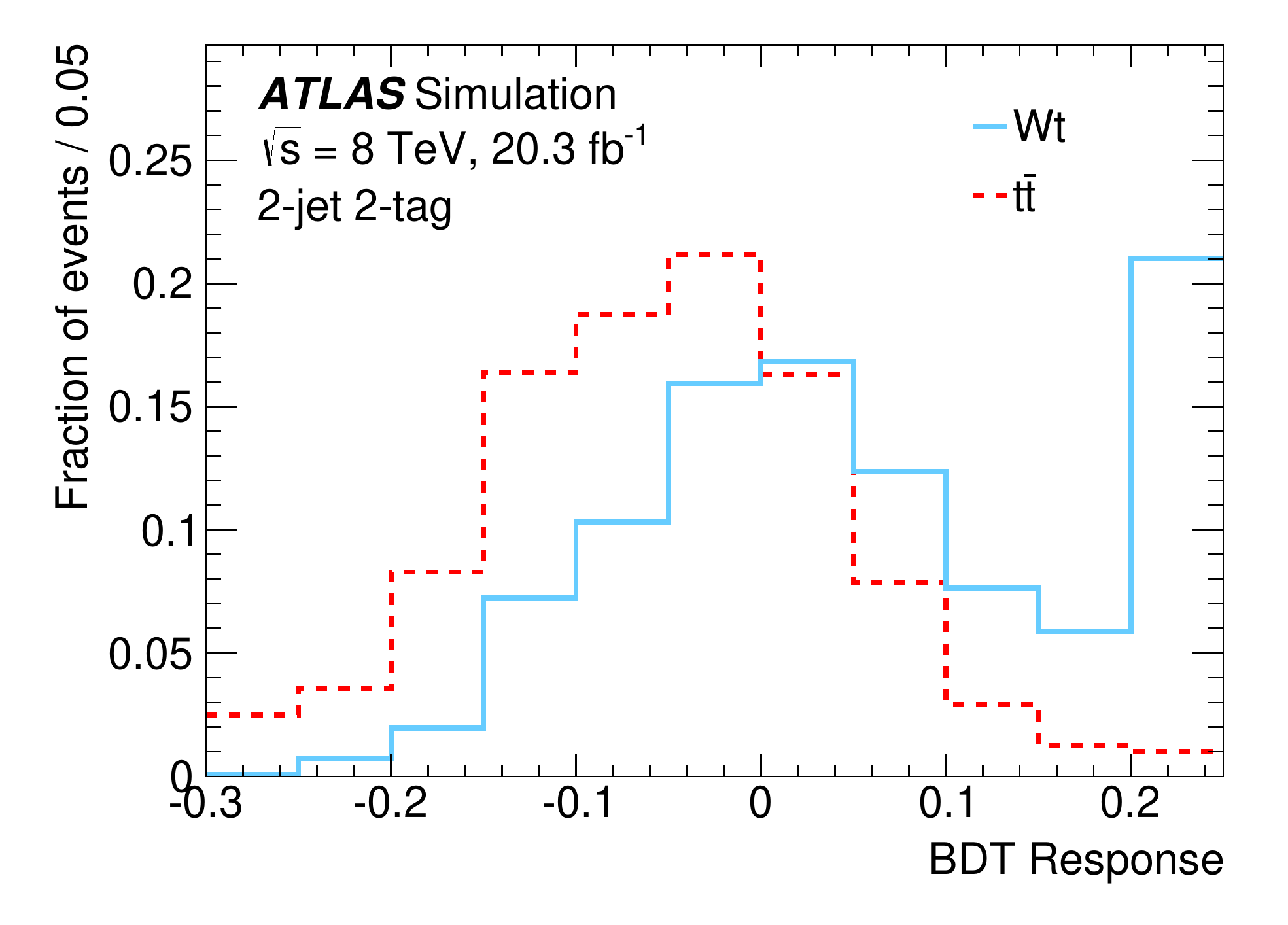}
}
\subfigure[]{\includegraphics[width=0.48\textwidth]{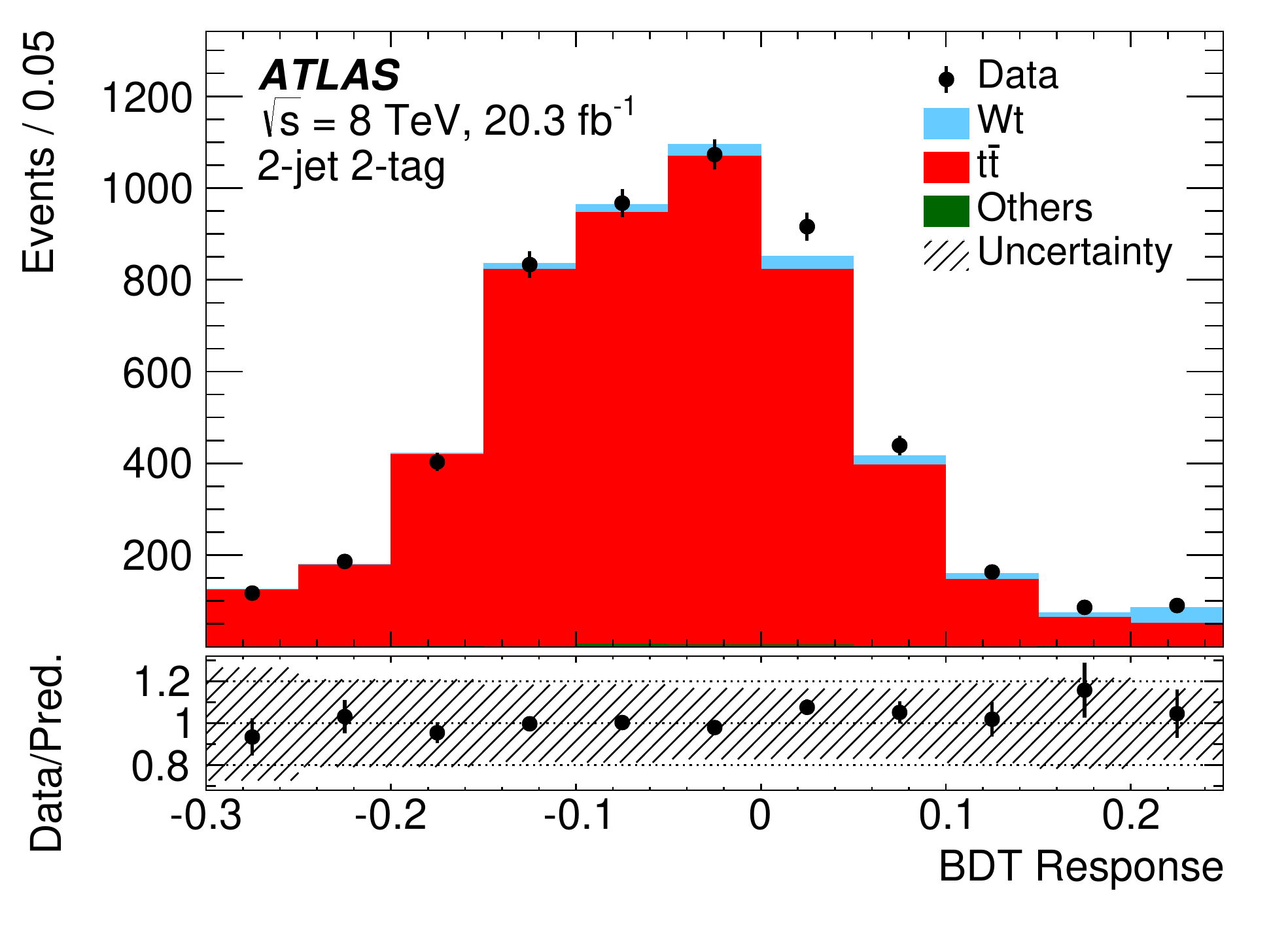}}
\caption{BDT response for (a, b) 1-jet 1-tag, (c, d) 2-jet 1-tag and (e, f) 2-jet 2-tag events.
Each contribution is normalised to unit area in (a, c, e) and to its expectation in (b, d, f).
The hatched area represents the sum in quadrature of the statistical and
systematic uncertainties.
The first bin includes the underflow and the last bin the overflow.
\label{fig:BDT_1jet}
}
  \end{center}
\end{figure}

The BDT response for the three regions is shown in Figure~\ref{fig:BDT_1jet}. The $Wt$ signal is
larger at positive BDT response values, while the $\ttbar$ background dominates for
negative BDT response values. The BDT range in each region is chosen to ensure sufficient simulation
statistics in each bin. The BDT separates
the signal from the background in all three regions, although even for high BDT response
values in the 1-jet 1-tag region, there remains a large expected background from
$\ttbar$ events.
The BDT responses from Figure~\ref{fig:BDT_1jet} are used in the profile likelihood fit swith this binning.

\clearpage
%-------------------------------------------------------------------------------
\section{Systematic uncertainties}
\label{sec:sys}
%-------------------------------------------------------------------------------

Systematic uncertainties affect the acceptance estimates for the signal and
background processes. Some of the systematic uncertainties also
affect the shape of the BDT response. 
Experimental sources of uncertainty arise from the modelling of
jets, leptons and $\etmiss$.

The impact of the uncertainty in the jet energy scale (JES) on the acceptance
and shape of the BDT response for $Wt$ and $\ttbar$ is evaluated in 22 uncorrelated components,
each of which can have a $\pt$ and $\eta$ dependence~\cite{PERF-2011-03,PERF-2012-01}.
The largest components are related to the modelling and the heavy-flavour correction, with
an acceptance uncertainty for $Wt$ and $\ttbar$ events of 1--2\%. The shape uncertainty is 
taken into account for the JES component with the largest impact on the
fit result (JES modelling component 1).
The jet energy resolution uncertainty is evaluated by smearing the energy of each jet in the
simulation and symmetrising the resulting change in acceptance and BDT response
shape~\cite{PERF-2011-04}. 
The resulting acceptance uncertainty for $Wt$ and $\ttbar$ events is 1--3\%, and the shape
uncertainty is taken into account.

The uncertainties in the modelling of the jet reconstruction and the jet vertex
fraction requirement are evaluated by randomly discarding jets according to the
difference in jet reconstruction efficiency between the data and MC simulation and by varying the
the jet vertex fraction requirement, respectively. These uncertainties have an impact on
the acceptance for $Wt$ and $\ttbar$ events of less than 1\%. They do not change the shape of the BDT response.

Further uncertainties arise from  the modelling of the  trigger, reconstruction, and
identification efficiencies for electrons~\cite{ATL-CONF-2014-032} and
muons~\cite{PERF-2014-05}, as well as from the modelling of the electron and muon energy
scale and resolution~\cite{PERF-2014-05,PERF-2013-05}. These have an
effect on the acceptance for $Wt$ and $\ttbar$ events of less than 1\%, except for the 
electron identification uncertainty, which has an acceptance
uncertainty for $Wt$ and $\ttbar$ of 2\%.
These uncertainties do not change the shape of the BDT response.

Uncertainties in the modelling of the $b$-tagging efficiency and mis-tag rates are estimated
from data~\cite{ATLAS-Perf-btag,ATLAS-CONF-2014-004}. These uncertainties depend on the
jet flavour and $\pt$, and for mis-tag rates also on jet $\eta$.
The uncertainty for $b$-jets is evaluated in six
components, with the largest component having an acceptance uncertainty for
$Wt$ and $\ttbar$ events of 1--4\%, depending on the analysis region~\cite{ATLAS-CONF-2014-004}.
The $b$-tag modelling uncertainties do not change the shape of the BDT response.

The variations in lepton and jet energies are propagated to the $\etmiss$ value.
This uncertainty has additional contributions from the modelling of the energy deposits which are not associated with any reconstructed object~\cite{PERF-2011-07}. Both an
energy scale and an energy resolution component are considered. The corresponding acceptance
uncertainty for $Wt$ and $\ttbar$ events is less than 0.3\%. The $\etmiss$ scale component also alters the shape of the BDT response.

\begin{figure}[htbp]
  \begin{center}
\subfigure[]{
  \includegraphics[width=0.48\textwidth]{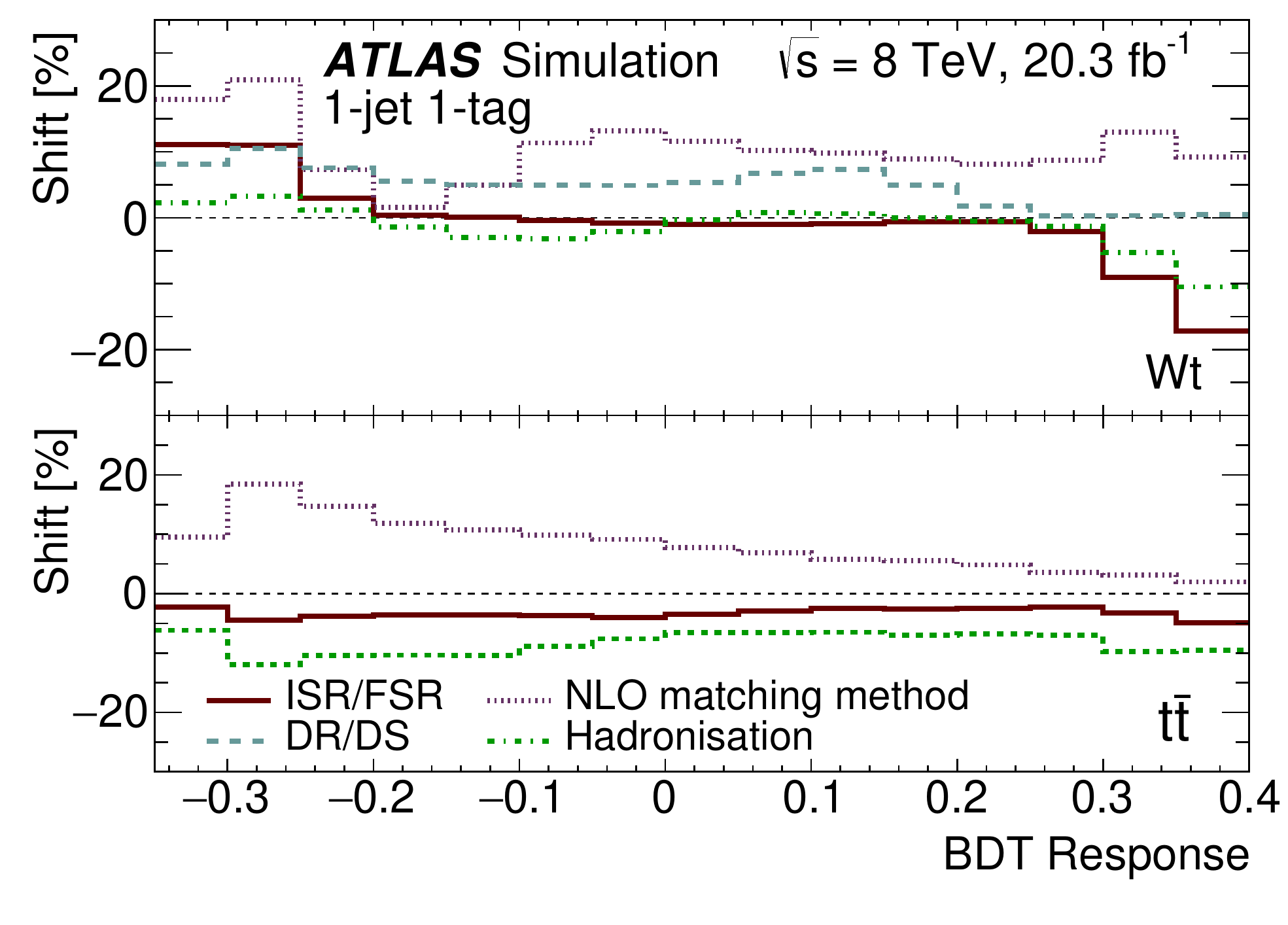}
 \label{fig:sysshapesa}
}
\subfigure[]{
  \includegraphics[width=0.48\textwidth]{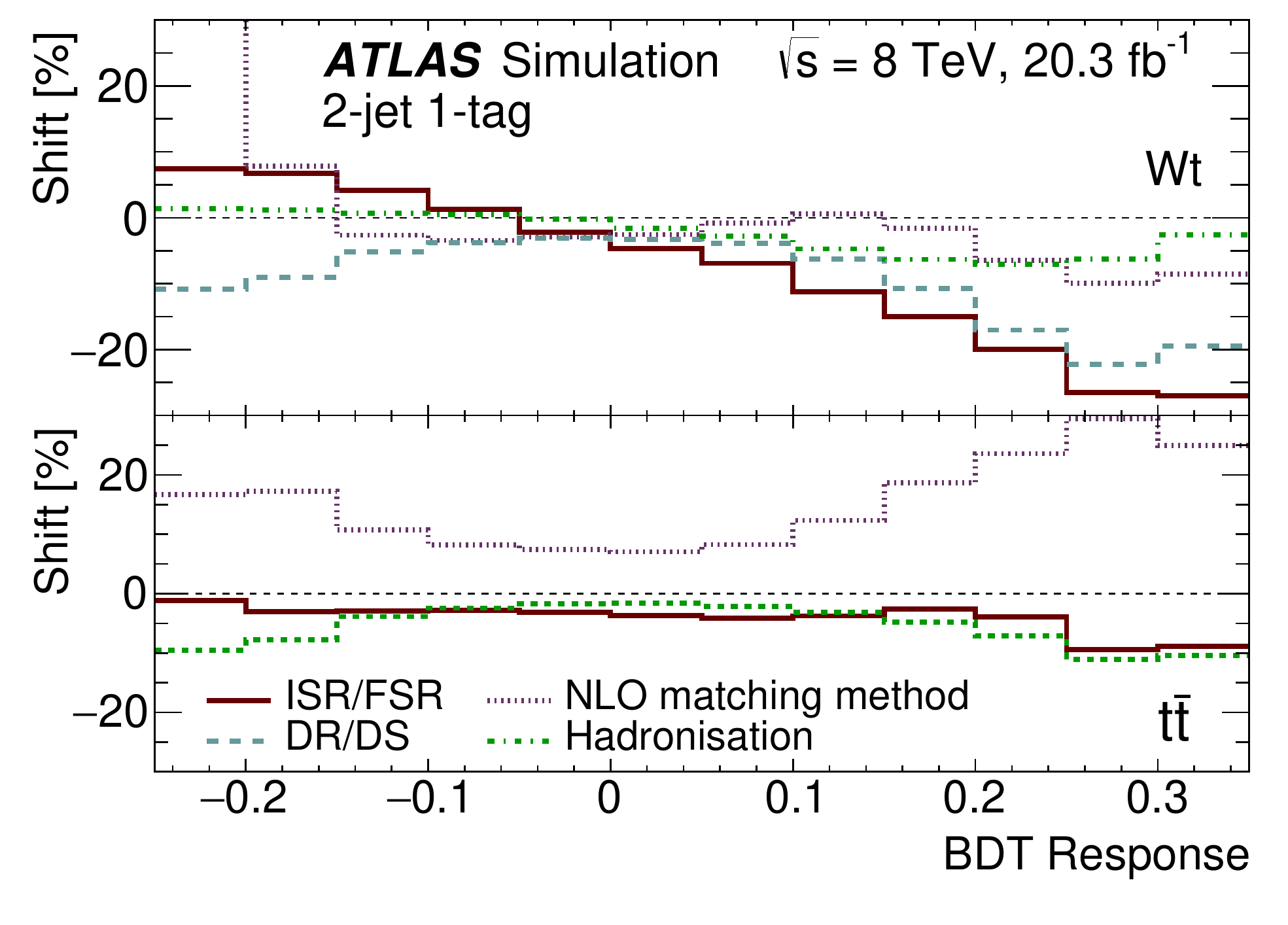}
 \label{fig:sysshapesb}
}
\subfigure[]{
  \includegraphics[width=0.48\textwidth]{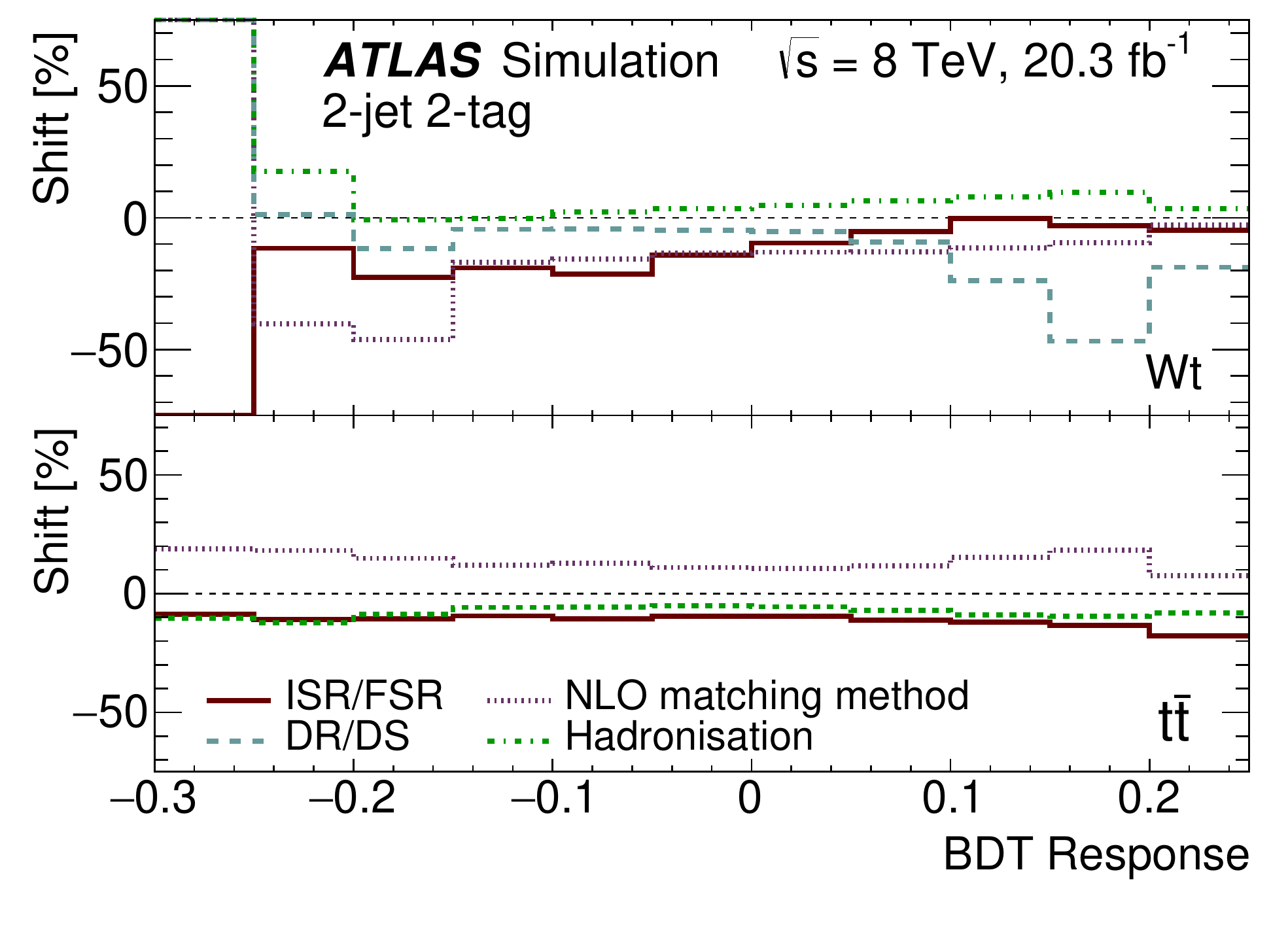}
 \label{fig:sysshapesc}
}
\caption{Relative shift of the BDT response associated with systematic variations of
 ISR/FSR, NLO matching method, DR/DS and hadronisation for (a) 1-jet 1-tag, (b) 2-jet 1-tag,
and (c) 2-jet 2-tag events. DR refers to the diagram-removal scheme, DS to the
diagram-subtraction scheme.
\label{fig:sysshapes}
}
  \end{center}
\end{figure}

Theoretical uncertainties are evaluated for the signal as well as the $\ttbar$
predictions. Figure~\ref{fig:sysshapes} shows the relative shift of the BDT response
associated with four of the theory modelling uncertainties.
The uncertainty on the $Wt$ signal and the $\ttbar$ background associated with initial-
and final-state radiation (ISR/FSR) is evaluated using {\scshape Powheg-Box}
interfaced to {\scshape Pythia}. The renormalisation scale associated with the strong
coupling $\alpha_\mathrm{S}$ is varied up and down by a factor of two 
in the matrix-element calculation and a
{\scshape Pythia} Perugia 2012 tune is used to create samples with increased and decreased levels of
radiation that are compatible with 7~\TeV\,ATLAS data~\cite{ATLAS:2012al}.
For $\ttbar$, the {\scshape hdamp} parameter of {\scshape Powheg-Box}~\cite{Alioli:2010xd},
which affects the amount of QCD radiation, is varied together with ISR/FSR. 
This uncertainty is treated as uncorrelated between $Wt$ and $\ttbar$ events.
Figure~\ref{fig:sysshapes} shows that this uncertainty has a large effect on the acceptance
and also alters the shape of the BDT response.

The uncertainty associated with the NLO matching method is evaluated by comparing {\scshape Powheg-Box} with {\scshape MC@NLO}, both interfaced to {\scshape Herwig}. 
Figure~\ref{fig:sysshapes} shows that this uncertainty has a dependence on the shape
of the BDT response.
For $Wt$ production, the largest impact of this uncertainty is
to shift events between the 1-jet 1-tag and 2-jet 2-tag regions. For $\ttbar$ events, the
impact of this uncertainty is on the acceptance, where it is 11--12\%.
This uncertainty is treated as correlated between $Wt$ and $\ttbar$ events.

The uncertainty associated with the modelling of the hadronisation and parton shower
is evaluated by comparing samples where {\scshape Powheg-Box} is interfaced with
{\scshape Pythia} to those where it is interfaced with {\scshape Herwig}.
This uncertainty alters the shape of the BDT response.

For the $Wt$ signal, the uncertainty associated with the scheme used to remove overlap with
$\ttbar$ is evaluated by comparing the two different schemes: the nominal sample, generated
with the DR scheme, is compared to a sample generated with the DS scheme. 
The relative shift of the BDT response is shown in Figure~\ref{fig:sysshapes}.
The relative shift of this uncertainty is
about 5\% in the signal region for 1-jet 1-tag events, and grows to large values in the
background-dominated region for 2-jet events, where its evaluation is limited by
simulation statistics and the predicted event yield is very small. 
This uncertainty alters the shape of the BDT response.

The evaluation of the PDF uncertainty follows the PDF4LHC
prescription~\cite{Botje:2011sn} using three different PDF sets ({\scshape CT10},
{\scshape MSTW2008nlo68cl}~\cite{Martin:2009iq} and {\scshape NNPDF2.3}~\cite{Ball:2013hta}).
The uncertainty on the acceptance for $Wt$ and $\ttbar$ events is evaluated in each of the three analysis regions.
The PDF uncertainty is considered correlated between $Wt$ and $\ttbar$ events,
except for $\ttbar$ 1-jet events, for which it is considered to be
uncorrelated. The PDF uncertainty components that affect the $\ttbar$ acceptance
in this region differ from the uncertainty components that affect the $\ttbar$ 
acceptance in the other regions~\cite{pdfstudy}.

The normalisation of the $\ttbar$ background has an uncertainty of
6\%~\cite{Czakon:2013goa,Czakon:2011xx}. The diboson background process has an uncertainty of
30\% for 1-jet events and 40\% for 2-jet events~\cite{Anastasiou:2003ds}, which is treated as
uncorrelated between different regions. The $Z$+jets and non-prompt or fake-lepton backgrounds
have normalisation uncertainties of 60\% to account for possible mismodelling of the jet
multiplicity and the acceptance of these small
backgrounds~\cite{Berger:2010zx,Berends:1990ax}. The $Z$+jets and non-prompt or fake-lepton
normalisation uncertainties are treated as uncorrelated between background sources and regions.

The uncertainty on the integrated luminosity is 2.8\%. It is
derived, following the same methodology as that detailed in
Ref.~\cite{DAPR-2011-01}, from a preliminary calibration of the
luminosity scale derived from beam-separation scans performed in
November 2012. The luminosity uncertainty enters in the extraction
of the cross-section as well as in the normalisation of the background processes that
are normalised to theory predictions. The statistical uncertainty due to the
finite size of the simulation samples is also taken into account.

%-------------------------------------------------------------------------------
\section{Results}
\label{sec:result}
%-------------------------------------------------------------------------------

\subsection{Measurement of the inclusive cross-section}
\label{subsec:xs}
A profile likelihood fit to the BDT classifier distributions is performed, using the RooStats software~\cite{Cranmer:2012sba,Moneta:2010pm}, in
order to determine the inclusive $Wt$ cross-section, utilising the 1-jet 1-tag, 2-jet 1-tag, and
2-jet 2-tag regions.  
The inclusion of the 2-jet regions provides additional
signal sensitivity and also helps to constrain the \ttbar background normalisation and
systematic uncertainties. 

The binned likelihood function is constructed as the product of Poisson probability terms
over all bins considered in the analysis. This likelihood depends on the signal-strength
parameter $\mu$, which is a multiplicative factor on the unconstrained $Wt$ yield prediction.
Nuisance parameters
(denoted $\theta$) are used to encode the effects of the various sources of systematic
uncertainty on the signal and background expectations. These nuisance parameters 
are implemented in the likelihood function with multiplicative Gaussian or log-normal constraints
with mean $\theta_0$ and standard deviation $\Delta \theta$.
The likelihood is then maximised with respect to the full set of $\mu$ and $\theta$
parameters. The values of these parameters after maximisation are referred to as
$\hat{\mu}$, $\hat{\theta}$, and $\Delta \hat{\theta}$.

The expected cross-section is obtained from a fit to the so-called Asimov dataset~\cite{Cowan:2010js}, with the signal and all backgrounds scaled to their predicted sizes~\cite{Kidonakis:2010ux}.
The expected measurement is $\hat{\mu}_{\textrm{exp}}=1.00_{-0.18}^{+0.17}$.
The observed result for the signal strength is 
$\hat{\mu}_{\textrm{obs}}=1.03_{-0.17}^{+0.16}$,
which corresponds to a measured cross-section of
$\SI[parse-numbers=false]{\twxs \twstat\, (\textrm{stat.}) \twsys\, (\textrm{syst.}) \twlumi\,(\textrm{lumi.})}{\pb}$.
Including systematic uncertainties, the observed (expected) significance of the signal compared to the background-only hypothesis
is $\sigobs$ ($\sigexp$) standard deviations,
obtained using an asymptotic approximation~\cite{Cowan:2010js}.

\begin{figure}[htbp]
\begin{center}
\includegraphics[width=0.7\textwidth]{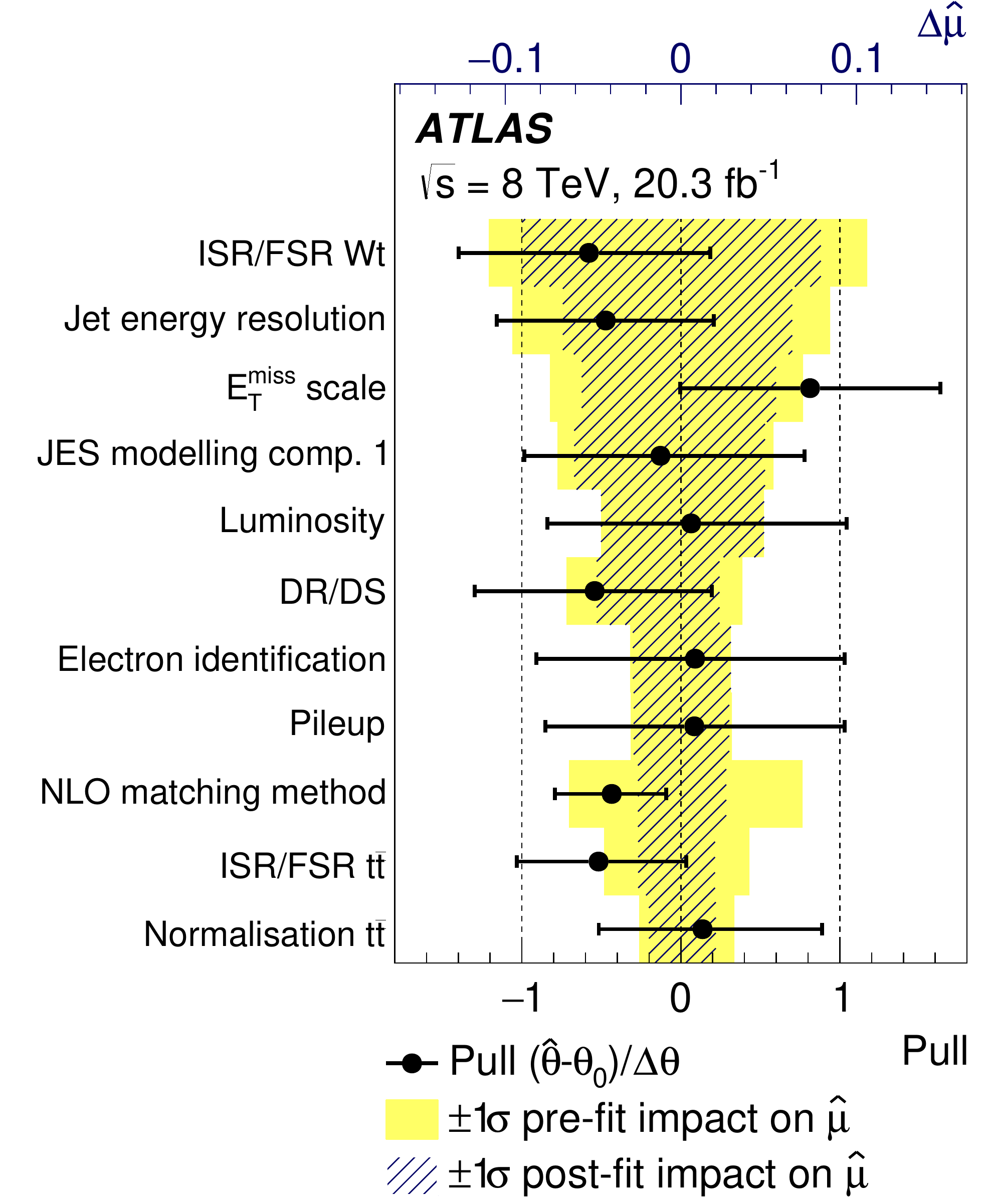}
\caption{Effect on the uncertainty on the fitted value of the signal strength $\hat{\mu}$
($\Delta\hat{\mu}$) and pull of the dominant nuisance parameters, ordered by their impact
on $\hat{\mu}$. The shaded and hashed areas refer to the
top axis: the shaded bands show the initial impact of that source of uncertainty on the
precision of $\hat{\mu}$; the hatched areas show the impact on the measurement of that source
of uncertainty, after the profile likelihood fit, at the $\pm 1 \sigma$ level.
The points and associated error bars show the pull of the nuisance parameters and
their uncertainties and refer to the bottom axis. A mean of zero and a width of 1 would imply no
constraint due to the profile likelihood fit. Only the 11 uncertainties with the largest
impact on $\hat{\mu}$ are shown.
\label{fig:pulls}
}
  \end{center}
\end{figure}

The post-fit (pre-fit) effect of each individual systematic uncertainty on $\hat{\mu}$ is
calculated by fixing the corresponding nuisance parameter at $\hat{\theta}+\Delta\hat{\theta}$ 
$\;(\hat{\theta}+\Delta\theta)$, and performing the fit again.
The difference between the default and the modified $\hat{\mu}$, $\Delta\hat{\mu}$,
represents the effect on $\hat{\mu}$ of this particular uncertainty. The pull
($(\hat{\theta}-\theta_0)/\Delta\theta$), and the pre-fit and post-fit impacts for the nuisance parameters
with the largest impact on $\hat{\mu}$ are shown in Figure~\ref{fig:pulls}. Since the total number of 
observed events in the 2-jet regions is about 14000, with a $Wt$ signal fraction of about 6\%, the
nuisance parameters that have a $\ttbar$ acceptance uncertainty of more than about 2\% can be
constrained in the fit. This applies to the jet energy resolution and $\ttbar$ normalisation uncertainties,
amongst others. The $\etmiss$ scale uncertainty has a shape dependence in the 1-jet 1-tag region
for $Wt$ and $\ttbar$, which results in the corresponding nuisance parameter being shifted but not much constrained.
The theory modelling uncertainties
due to ISR/FSR, DR/DS, and NLO matching method have large pre-fit and post-fit impacts. The nuisance
parameter for ISR/FSR $Wt$ is shifted and constrained in the fit due to its BDT response shape dependence,
shown in Figure~\ref{fig:sysshapes}. This uncertainty has the largest impact on $\hat{\mu}$, both
pre-fit and post-fit. The ISR/FSR $\ttbar$ uncertainty has a smaller post-fit impact
on $\hat{\mu}$ and is constrained due its acceptance and shape dependence.
In a test where the ISR/FSR uncertainty is considered to be correlated between $Wt$ and $\ttbar$ events, the expected uncertainty on $\hat{\mu}$ is reduced to $\pm 0.16$.
The nuisance parameter for the NLO matching method uncertainty is constrained by the
$\ttbar$ background because of the large acceptance component and
shape dependence of the NLO matching method uncertainty.

Table~\ref{tab:sysobs} summarises the contributions from the various
sources of systematic uncertainty to the uncertainties on the observed fit result. 
The total uncertainty in the table is the uncertainty obtained from
the full fit, and is therefore not identical to the sum in quadrature of the components, due to correlations that the fit induces between the uncertainties. The largest contributions
to the cross-section uncertainty are from the modelling of ISR/FSR and from the jet energy
resolution and scale.

\begin{table}[!h!tp]
\begin{center}
\caption{Summary of the relative uncertainties on the $Wt$ cross-section measurement.
Detector uncertainties are grouped into categories. All sources of uncertainty within 
a category are added in quadrature to obtain the category uncertainty.}
\label{tab:sysobs}
\begin{tabular}{|l|r|}
  \hline
Uncertainty & Impact on $\hat{\mu}$ [\%]\\[0.15cm]
\hline
Statistical       & $\pm 5.8$ \\[0.15cm]
Luminosity        & $\pm 4.7$ \\[0.15cm]
Theory modelling   & \\[0.15cm]
~~ISR/FSR         &  ${}^{+8.2}_{-9.4}$ \\[0.15cm]
~~Hadronisation   &  $\pm 1.7$ \\[0.15cm]
~~NLO matching method&  $\pm 2.5$  \\[0.15cm]
~~PDF             &  $\pm 0.6$ \\[0.15cm]
~~DR/DS           & ${}^{+2.2}_{-4.8}$ \\[0.15cm]
Detector    & \\[0.15cm]
~~Jet       &  ${}^{+9.0}_{-9.9}$ \\[0.15cm]
~~Lepton    &  $\pm 3.0$ \\[0.15cm]
~~$\etmiss$ &  $\pm 5.5$ \\[0.15cm]
~~$b$-tag   &  $\pm 1.0$ \\[0.15cm]
Background norm. & ${}^{+2.9}_{-2.6}$ \\[0.15cm]
\hline
 & \\[-0.35cm]
Total       & ${}^{+16}_{-17}$ \\[0.15cm]
  \hline
\end{tabular}
\end{center}
\end{table}

\begin{figure}[!h!tbp]
\begin{center}
\subfigure[]{
\includegraphics[width=0.48\textwidth]{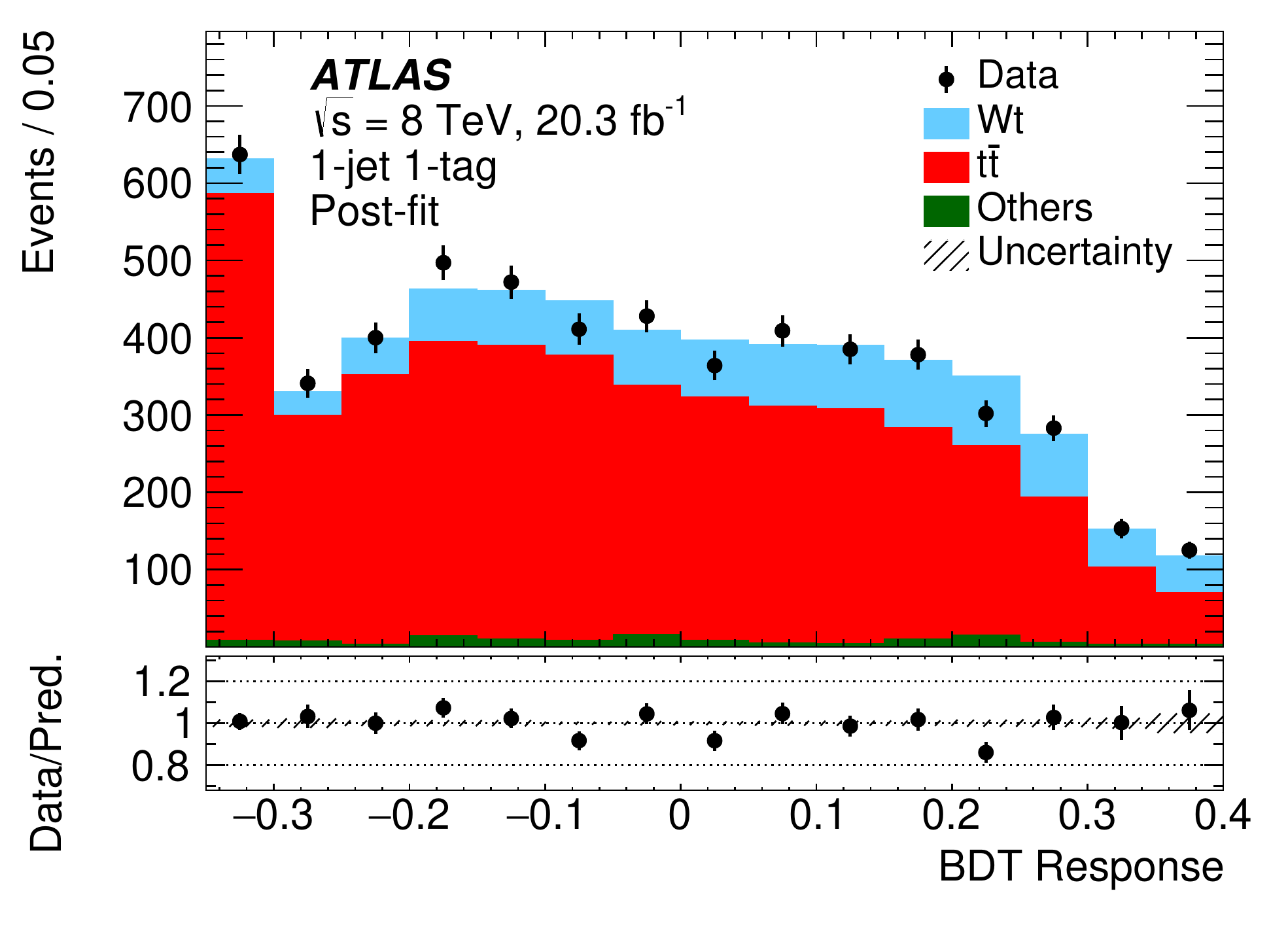}
}
\subfigure[]{
\includegraphics[width=0.48\textwidth]{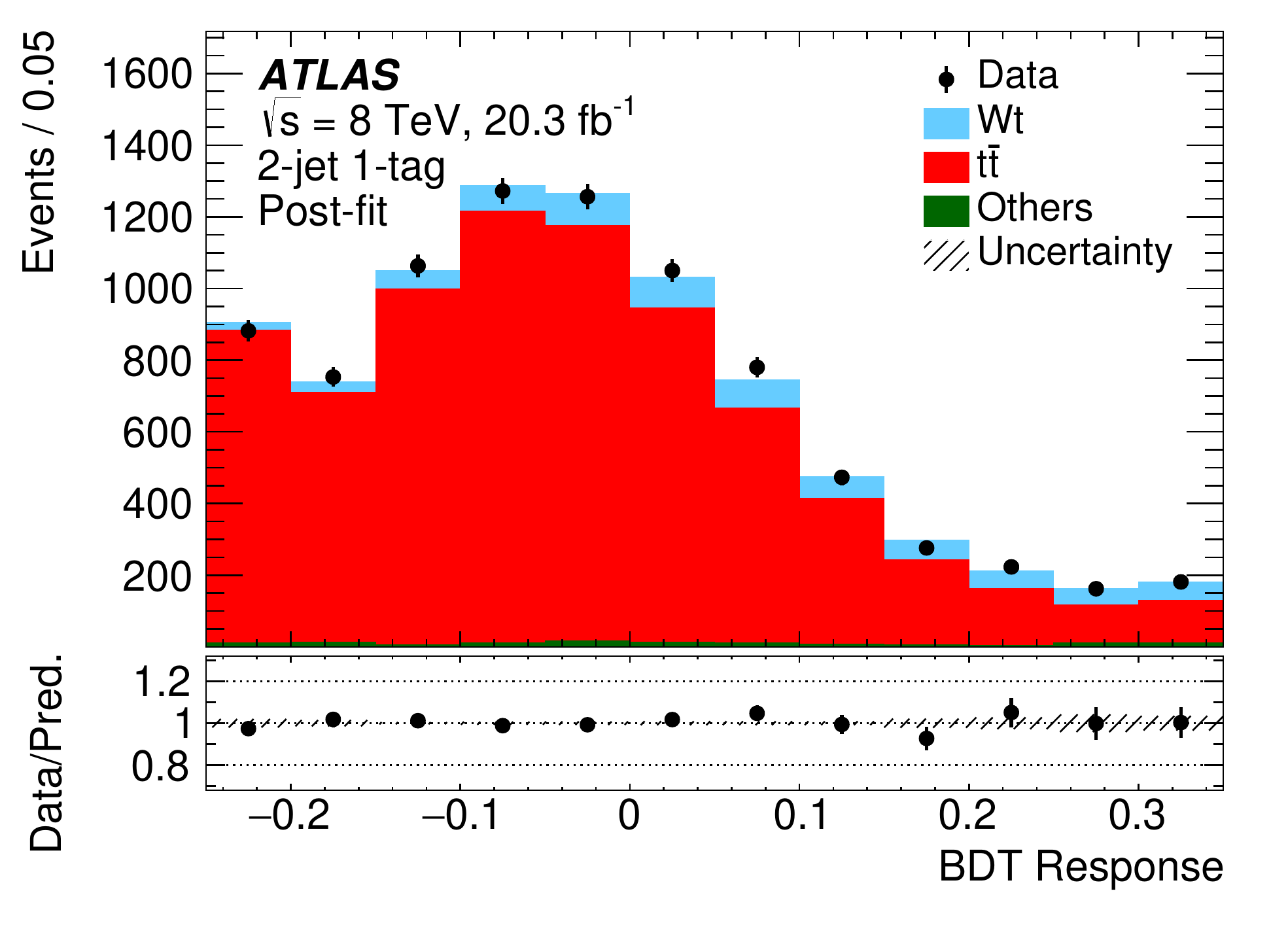}
}
\subfigure[]{
\includegraphics[width=0.48\textwidth]{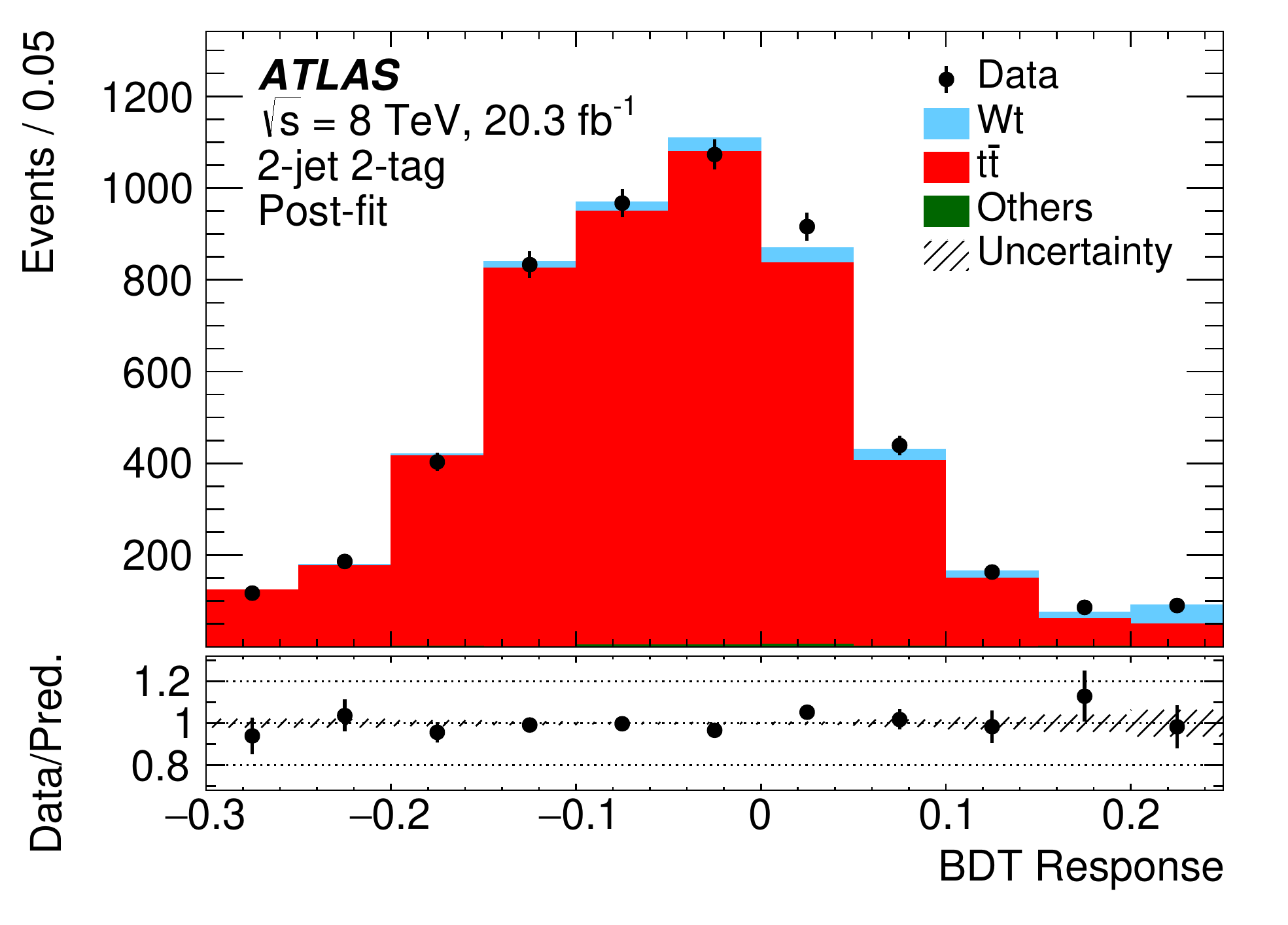}
}
\caption{Distribution of the post-fit BDT response for (a) 1-jet 1-tag, (b) 2-jet 1-tag,
and (c) 2-jet 2-tag events. The signal, backgrounds and uncertainties are scaled
to the fit result.
The first bin includes the underflow and the last bin the overflow.
\label{fig:bdtobs}
}
  \end{center}
\end{figure}

The BDT response for each region is shown normalised to the fit result in Figure~\ref{fig:bdtobs}. The dependence of the cross-section on the top-quark mass is evaluated using $Wt$ and $\ttbar$
simulation samples with various top-quark masses.
The cross-section depends linearly on the top-quark mass due to changes in acceptance,
with a slope of 1.11~pb/\GeV.

\subsection{Constraints on $|f_{\textrm{LV}}V_{tb}|$  and  $|V_{tb}|$}
\label{subsec:vtb}
The inclusive cross-section measurement provides a direct determination of the magnitude of
the CKM matrix element $V_{tb}$.
The ratio of the measured cross-section to the theoretical prediction is equal to
$|f_{\textrm{LV}}V_{tb}|^2$, where the form factor $f_{\textrm{LV}}$ could be modified
by new physics or radiative corrections through anomalous coupling contributions,
for example those in Refs.~\cite{AguilarSaavedra:2008zc,Kane:1991bg,Rizzo:1995uv}. The $Wt$ production and
top-quark decays through $|V_{ts}|$ and $|V_{td}|$ are assumed to be
small. A lower limit on $|V_{tb}|$ is obtained for $f_{\textrm{LV}}=1$ as in the SM,
without assuming CKM unitarity~\cite{Alwall:2006bx,Cao:2015qta}. 
An additional systematic uncertainty due to a variation of the top-quark mass
by 1~\GeV\,is included in the $V_{tb}$ extraction.
The uncertainties on the theoretical cross-section due to the
variation of the renormalisation and factorisation scale (0.6~pb),
the PDF uncertainty (1.4~pb), and the beam-energy
uncertainty~\cite{Wenninger:1546734} (0.38~pb) are also accounted for.

The value for $|f_{\textrm{LV}}V_{tb}|$ is extracted from the $|f_{\textrm{LV}}V_{tb}|^{2}$ likelihood,
which is assumed to be Gaussian. The lower limit on $|V_{tb}|^2$ corresponds to 95\% of the integral
of this likelihood, setting $f_{\textrm{LV}}=1$ and starting at~1.
The measured value of $|f_{\textrm{LV}}V_{tb}|$ is $\vtboval$, and the corresponding lower limit on  $|V_{tb}|$ at the 95\% confidence level is $\vtbolim$.

%-------------------------------------------------------------------------------
\section{Cross-section measurement inside a fiducial acceptance}
\label{sec:fid}
%-------------------------------------------------------------------------------

The cross-section for the production of events containing a top quark and a $W$~boson
is measured in a fiducial region to allow a more robust comparison to the theoretical prediction without extrapolating to regions outside of the detector acceptance.
The fiducial measurement reduces the sensitivity of the cross-section to theory modelling
uncertainties. The measurement can also be compared to particle-level predictions for the
inclusive $WWb$ and $WWbb$ processes at NLO, once those calculations become
available~\cite{Cascioli:2013wga,Heinrich:2013qaa}.
The fiducial acceptance requires two leptons and exactly one $b$-jet at the particle level.
This encompasses not only $Wt$ production but also $\ttbar$ production where one of
the $b$-quarks from the top-quark decays is not in the particle-level acceptance.
The fiducial cross-section is measured by fitting the sum of the $Wt$ and $\ttbar$ contributions
to data in the 1-jet 1-tag region. Control regions are not used in the fit.

\subsection{Fiducial selection}
\label{sec:fidsel}

The definition of the fiducial acceptance is based on MC simulation and uses particle-level physics
objects constructed of stable particles with a mean lifetime $\tau>0.3 \times 10^{-10}$~s. Electrons and muons are required to originate from $W$-boson decays, either directly or via leptonically decaying $\tau$ leptons. The $\pt$ of each of the leptons is corrected by adding the energy and momentum of photons inside a cone of size $\Delta R=0.1$ around the lepton direction. Electrons and muons are required to have $\pt>25$~\GeV\,and $|\eta|<2.5$. Jets are clustered from particles using the anti-$k_t$ algorithm with radius parameter $R=0.4$. Neutrinos, electrons and muons from $W$-boson decays as well as particles resulting from pileup are excluded from jet clustering. Particles from the underlying event are included. The particle-level jets are required to have $\pt>20$~\GeV\,and $|\eta|<2.5$ and are matched with nearby $b$-hadrons with a $\pt$ of at least 5~\GeV\,using the ghost tagging
method~\cite{Schaetzel:2013vka}. Jets within $\Delta R=0.2$ of the nearest electron are removed from the list.
Following that, electrons and muons within $\Delta R=0.4$ of the nearest jet are removed.
Missing transverse momentum is calculated using neutrinos from $W$-boson decays.
The $Wt$ and $\ttbar$ events pass the fiducial selection if they have exactly two leptons,
exactly one $b$-jet and $\etmiss>20$~\GeV. 
The numbers of simulated $Wt$ and $\ttbar$ events passing this fiducial selection are shown in
Table~\ref{tab:fidYield}, and $Wt$ production contributes 26\% of these particle-level events.

Simulated $Wt$ and $\ttbar$ events that satisfy the detector-level selection criteria are separated into two categories: in-fiducial (satisfying the fiducial selection criteria) and out-of-fiducial (the rest).
Table~\ref{tab:fidYield} shows the number of events for $Wt$ and $\ttbar$ in each category. The $Wt$ contribution is 25\% of the in-fiducial events, but only 10\% of the out-of-fiducial events.
The out-of-fiducial events that pass the detector-level selection typically have two or more
particle-level jets, only one of which is also reconstructed at the detector level.
Thus the $\ttbar$ contribution to the out-of-fiducial events is larger.

\begin{table}[!h!tp]
\begin{center}
\caption{Number of expected events at the particle-level and for the detector-level selection
for $Wt$ and $\ttbar$. The uncertainty for the particle-level includes ISR/FSR, NLO matching
method, and for $Wt$ also hadronisation, all added in quadrature.
The uncertainty for the detector-level selection includes all sources of uncertainty, added in quadrature.
}
\label{tab:fidYield}
\begin{tabular}{|l|
 S[table-figures-uncertainty=1]|
 S[table-figures-uncertainty=1]|
 S[table-figures-uncertainty=1]|
}
  \hline
                 & {Particle-level} & \multicolumn{2}{c|}{Detector-level selection} \\
Process          & {selection}      & {in-fiducial} & {out-of-fiducial}\\
\hline \hline
$Wt$             &  4200 \pm 100  &  810 \pm 160 &  230 \pm 40 \\ 
$\ttbar$         & 12000 \pm 2000 & 2400 \pm 500 & 2100 \pm 400 \\ 
\hline
\end{tabular}
\end{center}
\end{table}

\subsection{Systematic uncertainties}
\label{sec:fidsys}

The sources of systematic uncertainty in the inclusive cross-section measurement are also considered for the fiducial measurement. The object reconstruction and background-normalisation uncertainties also apply in this measurement (except the $\ttbar$ normalisation uncertainty, as discussed below).
For in-fiducial events, a variation in the theory modelling uncertainties (DR/DS, ISR/FSR, hadronisation, NLO matching method, and PDF) changes the detector-level and fiducial acceptances in the same direction, which reduces the impact of these uncertainties. Since this does not affect
out-of-fiducial events, these theory modelling uncertainties are treated as uncorrelated between in- and out-of-fiducial events.

An additional uncertainty accounts for the relative fractions of $Wt$ and $\ttbar$ due to the uncertainty on the theoretical predictions. The fraction of each type of signal is allowed to vary within their theoretical predictions, keeping the sum constant. 

\subsection{Results}
\label{sec:fidres}

The fiducial cross-section is measured in a profile likelihood fit to data in the 1-jet 1-tag
region. In-fiducial and out-of-fiducial $Wt$ and $\ttbar$ events are scaled by the same
cross-section scale factor $\mu_{\textrm{fid}}$ in the fit.
The measured fiducial cross-section for $Wt$ and $\ttbar$ production is
$\SI[parse-numbers=false]{\fidxs \fidstat\, (\textrm{stat.}) \fidsys\, (\textrm{syst.}) \fidlumi\,(\textrm{lumi.})}{\pb}$,
which corresponds to a total uncertainty of 8\%. The expected uncertainty is also 8\%.
The impact of the systematic uncertainties on this measurement is summarised
in Table~\ref{tab:sysfid}. The relative uncertainties are smaller in the
fiducial measurement than in the inclusive measurement (cf. Table~\ref{tab:sysobs})
because both $Wt$ and $\ttbar$ events are considered signal and because of the definition of
the fiducial acceptance. The only exception is the $b$-tag uncertainty, which is larger
in the fiducial measurement because only 1-jet 1-tag events are used in the fit.

\begin{table}[!h!tp]
\begin{center}
\caption{Summary of the uncertainties on the observed fit result for the fiducial
cross-section. Detector uncertainties are grouped into categories. All sources of
uncertainty within a category are added in quadrature to obtain the category uncertainty.}
\label{tab:sysfid}
\begin{tabular}{|l|r|}
  \hline
Uncertainty & Impact on $\hat{\mu}_{\textrm{fid}}$ [\%]\\ [0.15cm]
\hline
Statistical       & 1.0 \\
Luminosity        & 3.1 \\
Theory modelling   & \\
~~ISR/FSR         &  4.2 \\
~~Hadronisation   &  0.8 \\
~~NLO matching method &  0.7 \\
~~PDF             &  <0.1 \\
~~Ratio $Wt/\ttbar$& 2.2 \\
~~DR/DS           & 0.1 \\
Detector    & \\
~~Jet       &  5.2 \\
~~Lepton    &  2.3 \\
~~$\etmiss$ &  0.2 \\
~~$b$-tag   &  2.3 \\
Background norm. & <0.1 \\
\hline
Total       & 8.2 \\
  \hline
\end{tabular}
\end{center}
\end{table}

\begin{figure}[!h!tbp]
\begin{center}
\includegraphics[width=0.8\textwidth]{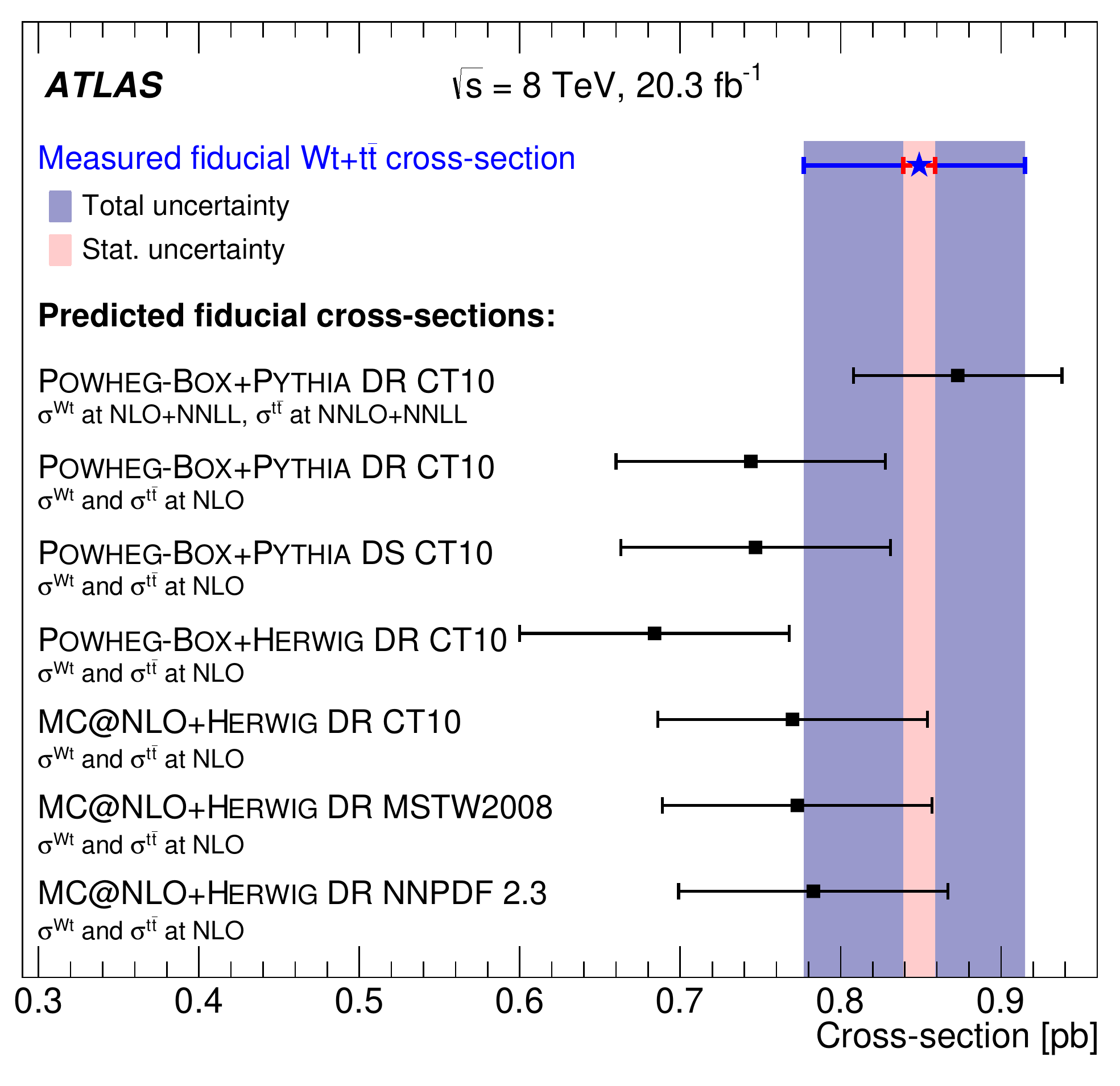}
\caption{Comparison of the measured fiducial cross-section to theoretical predictions
in a fiducial acceptance requiring two leptons with $\pt>25$~\GeV\,and $|\eta|<2.5$,
one jet with $\pt>20$~\GeV\,and $|\eta|<2.5$, and $\etmiss>20$~\GeV.
The predictions are computed at NLO accuracy for the
fiducial acceptance and the inclusive cross-section, except for the top line, for
which the inclusive cross-sections for $Wt$ and $\ttbar$ are computed at NLO+NNLL and NNLO+NNLL accuracy,
respectively.
\label{fig:fid}
}
  \end{center}
\end{figure}

The measured fiducial cross-section is compared to theoretical predictions for the
sum of the fiducial $Wt$ and $\ttbar$ cross-sections in Figure~\ref{fig:fid}.
The uncertainty on the theory predictions accounts for scale and PDF contributions.
The MSTW2008 and NNPDF2.3 predictions are obtained by re-weighting the 
simulated {\scshape{Mc@nlo}} sample.
The uppermost result for the predicted fiducial cross-section is based on the fiducial
acceptances and the sample normalisation utilised in this analysis.
The fiducial acceptances are computed
from the nominal {\scshape{Powheg-Box+Pythia}} samples. The $Wt$ and $\ttbar$ cross-sections are
normalised to their NLO+NNLL and NNLO+NNLL predictions, respectively. The other results utilise
the theoretical cross-sections as computed by the respective generator.

% All figures and tables should appear before the summary and conclusion
% The package placeins provides the macro \FloatBarrier to achieve this
\FloatBarrier
%-------------------------------------------------------------------------------
\section{Conclusion}
\label{sec:conclusion}
%-------------------------------------------------------------------------------

The inclusive cross-section for the production of a single top quark in association with a
$W$~boson has been measured in proton-proton collisions at a centre-of-mass energy of 8~\TeV,
using dilepton events from $20.3 \rm \ fb^{-1}$ of data recorded by the ATLAS detector at the LHC.
$Wt$ production is observed with a significance of
\SI[parse-numbers=false]{\sigobs}{$\sigma$}. 
The measured cross-section is
$$\SI[parse-numbers=false]{\twxs \twstat\, (\textrm{stat.}) \twsys\, (\textrm{syst.}) \twlumi\,(\textrm{lumi.})}{\pb}\,,$$
in agreement with the NLO+NNLL expectation.
The measured cross-section is used to extract a direct measurement of the left-handed form factor
times the CKM
matrix element $|f_{\textrm{LV}}V_{tb}|$ of \SI[parse-numbers=false]{\vtboval}{}.
The lower limit on $|V_{tb}|$ is \SI[parse-numbers=false]{\vtbolim} at the 95\% CL,
without assuming unitarity of the CKM matrix.
The cross-section for the production of a $W$~boson and a top quark (including $Wt$ and $\ttbar$)
has also been measured in a fiducial acceptance requiring two leptons with $\pt>25$~\GeV\,and $|\eta|<2.5$,
one jet with $\pt>20$~\GeV\,and $|\eta|<2.5$, and $\etmiss>20$~\GeV. 
The fiducial cross-section is 
$$\SI[parse-numbers=false]{\fidxs \fidstat\, (\textrm{stat.}) \fidsys\, (\textrm{syst.}) \fidlumi\,(\textrm{lumi.})}{\pb}\,.$$

%-------------------------------------------------------------------------------
\section*{Acknowledgements}
%-------------------------------------------------------------------------------

% Acknowledgements for papers with collision data
% Version 15-Sep-2015

% Standard acknowledgements start here
%----------------------------------------------
We thank CERN for the very successful operation of the LHC, as well as the
support staff from our institutions without whom ATLAS could not be
operated efficiently.

We acknowledge the support of ANPCyT, Argentina; YerPhI, Armenia; ARC, Australia; BMWFW and FWF, Austria; ANAS, Azerbaijan; SSTC, Belarus; CNPq and FAPESP, Brazil; NSERC, NRC and CFI, Canada; CERN; CONICYT, Chile; CAS, MOST and NSFC, China; COLCIENCIAS, Colombia; MSMT CR, MPO CR and VSC CR, Czech Republic; DNRF, DNSRC and Lundbeck Foundation, Denmark; IN2P3-CNRS, CEA-DSM/IRFU, France; GNSF, Georgia; BMBF, HGF, and MPG, Germany; GSRT, Greece; RGC, Hong Kong SAR, China; ISF, I-CORE and Benoziyo Center, Israel; INFN, Italy; MEXT and JSPS, Japan; CNRST, Morocco; FOM and NWO, Netherlands; RCN, Norway; MNiSW and NCN, Poland; FCT, Portugal; MNE/IFA, Romania; MES of Russia and NRC KI, Russian Federation; JINR; MESTD, Serbia; MSSR, Slovakia; ARRS and MIZ\v{S}, Slovenia; DST/NRF, South Africa; MINECO, Spain; SRC and Wallenberg Foundation, Sweden; SERI, SNSF and Cantons of Bern and Geneva, Switzerland; MOST, Taiwan; TAEK, Turkey; STFC, United Kingdom; DOE and NSF, United States of America. In addition, individual groups and members have received support from BCKDF, the Canada Council, CANARIE, CRC, Compute Canada, FQRNT, and the Ontario Innovation Trust, Canada; EPLANET, ERC, FP7, Horizon 2020 and Marie Skłodowska-Curie Actions, European Union; Investissements d'Avenir Labex and Idex, ANR, Region Auvergne and Fondation Partager le Savoir, France; DFG and AvH Foundation, Germany; Herakleitos, Thales and Aristeia programmes co-financed by EU-ESF and the Greek NSRF; BSF, GIF and Minerva, Israel; BRF, Norway; the Royal Society and Leverhulme Trust, United Kingdom.

The crucial computing support from all WLCG partners is acknowledged
gratefully, in particular from CERN and the ATLAS Tier-1 facilities at
TRIUMF (Canada), NDGF (Denmark, Norway, Sweden), CC-IN2P3 (France),
KIT/GridKA (Germany), INFN-CNAF (Italy), NL-T1 (Netherlands), PIC (Spain),
ASGC (Taiwan), RAL (UK) and BNL (USA) and in the Tier-2 facilities
worldwide.
%----------------------------------------------

%-------------------------------------------------------------------------------
\clearpage
\appendix

\printbibliography

\newpage
% ATLAS Collaboration author list
% Data extracted on 28-Aug-2015 for paper reference TOPQ-2012-20
\begin{flushleft}
{\Large The ATLAS Collaboration}

\bigskip

G.~Aad$^{\rm 85}$,
B.~Abbott$^{\rm 113}$,
J.~Abdallah$^{\rm 151}$,
O.~Abdinov$^{\rm 11}$,
R.~Aben$^{\rm 107}$,
M.~Abolins$^{\rm 90}$,
O.S.~AbouZeid$^{\rm 158}$,
H.~Abramowicz$^{\rm 153}$,
H.~Abreu$^{\rm 152}$,
R.~Abreu$^{\rm 116}$
% ...
\bigskip
\\
% ...
$^{11}$ Institute of Physics, Azerbaijan Academy of Sciences, Baku, Azerbaijan\\
% ...
$^{85}$ CPPM, Aix-Marseille Universit{\'e} and CNRS/IN2P3, Marseille, France\\
% ...
$^{90}$ Department of Physics and Astronomy, Michigan State University, East Lansing MI, United States of America\\
% ...
$^{107}$ Nikhef National Institute for Subatomic Physics and University of Amsterdam, Amsterdam, Netherlands\\
% ...
$^{113}$ Homer L. Dodge Department of Physics and Astronomy, University of Oklahoma, Norman OK, United States of America\\
% ...
$^{116}$ Center for High Energy Physics, University of Oregon, Eugene OR, United States of America\\
% ...
$^{151}$ Institute of Physics, Academia Sinica, Taipei, Taiwan\\
$^{152}$ Department of Physics, Technion: Israel Institute of Technology, Haifa, Israel\\
$^{153}$ Raymond and Beverly Sackler School of Physics and Astronomy, Tel Aviv University, Tel Aviv, Israel\\
% ...
$^{158}$ Department of Physics, University of Toronto, Toronto ON, Canada\\
% ...
\end{flushleft}

% Created with xml2latex.py

\end{document}